\documentclass[11pt]{article}

\usepackage[margin=1in]{geometry}
\usepackage{setspace}
\usepackage{lmodern}
\usepackage[T1]{fontenc}
\usepackage{microtype}
\usepackage{amsmath,amssymb,amsfonts}
\usepackage{graphicx}
\usepackage{booktabs}
\usepackage[round]{natbib}
\usepackage[hidelinks]{hyperref}
\usepackage{xurl}

\graphicspath{{./}{figures/}{../figures/}}

\newcommand{\backmatter}{}
\newcommand{\bmhead}[1]{\subsection*{#1}}

\hypersetup{pdftitle={What exam scores can and cannot prove about unauthorized AI assistance},
            pdfauthor={Chad M. Topaz and Utsav Bahl}}
\title{What exam scores can and cannot prove about unauthorized AI
assistance: Evidence from a highly public classroom episode}

\author{Chad M.\ Topaz\textsuperscript{1,2,3,*} and Utsav Bahl\textsuperscript{4}\\[8pt]
\small \textsuperscript{1}Williams College, Williamstown, MA 01267, USA\\[-2pt]
\small \textsuperscript{2}QSIDE Institute, Williamstown, MA 01267, USA\\[-2pt]
\small \textsuperscript{3}University of Colorado, Boulder, CO 80309, USA\\[-2pt]
\small \textsuperscript{4}University of Cambridge, Cambridge CB2 1TN, UK\\[6pt]
\small \textsuperscript{*}Corresponding author: cmt6@williams.edu}
\date{\today}

\begin{document}
\maketitle

\begin{abstract}
\noindent
In spring 2026, an economics professor at Brown University gave a take-home midterm and, after unusually high scores, made the final exam proctored. Among the 59 students who completed the course, average scores fell from 95.7 out of 100 to 48.8. The instructor attributed the drop to unauthorized use of generative AI on the midterm; others proposed test anxiety, a harder final, student withdrawals, and regression to the mean. The publicly released scores, one midterm--final pair per student, show two striking patterns: the correlation between a student's two scores is only 0.06, and individual changes range from a 4-point gain to a 100-point loss. If both exams measured the same proficiency, students who scored higher on the midterm should generally have scored higher on the final, and most scores should have fallen by similar amounts. Permutation tests find no statistically detectable association between students' midterm and final scores, yet would detect association of the strength the alternative explanations predict more than 90 percent of the time. Once model complexity is accounted for, a model in which a student's midterm carries no information about that student's final describes the data as well as any model that permits an association. Neither result rules out a weak association, but together they show that the data do not require any. In simulated classes where each student's two scores remain linked through that student's own proficiency, as the alternative explanations imply, the two patterns almost never appear together; they appear together regularly only when that link is nearly severed. More than one mechanism could have severed it: midterm answers that were not the students' own work would have done so, and so would a final that tested substantially different knowledge and skills, with no assistance involved. The scores cannot distinguish these possibilities or identify which students, if any, used unauthorized assistance. Our study offers a framework for evaluating statistical evidence in future disputes over generative AI use in academic assessment.
\end{abstract}


\bigskip

\section{Introduction}\label{sec:intro}

In the spring of 2026, Roberto Serrano, a professor of economics at Brown University, administered a take-home, closed-book midterm to his class of 86 students and received back a set of papers averaging 96 out of 100, with roughly 40 perfect scores \citep{serrano2026fp, gioino2026}. His previous midterms, which he describes as easier, had averaged between 65 and 80 \citep{serrano2026fp}. He then announced that the final examination would instead be proctored in person, and that he would void the midterm if the two grade distributions failed to match. Following the announcement, 27 of the 86 students either dropped the course or skipped the final, and the 59 students who remained averaged approximately 49, below any final-exam mean he reports having recorded in the course \citep{ihe2026, goldstein2026}. He concluded that at least 50 students had made unauthorized use of generative artificial intelligence (AI) on the midterm, and said so publicly \citep{serrano2026fp, goldstein2026}.

The episode became national news, and public reaction divided between two interpretations. On one, the 47-point collapse was decisive evidence of mass cheating. On the other, advanced mainly in discussion forums, the collapse could be explained without any unauthorized assistance (hereafter, \emph{assistance-free} explanations). For instance, students perform better with more time and less anxiety; the final may have been harder; the class had tripled in size when the take-home format was announced, changing its composition; and a score near the top of the scale has nowhere to go but down.

Whether these explanations can account for the data, singly or in combination, is a quantitative question. We answer it by giving each explanation an explicit numerical form, sized from the education-measurement literature where that literature offers a bound and set generously where it does not, and then asking whether classes simulated under those explanations produce score patterns like the ones in the released data.
The analysis is possible because the student-level data are public. Press coverage of the episode published the anonymized midterm and final scores of all 59 students who completed the course, reporting them as the data the professor submitted to Brown's Standing Committee on the Academic Code \citep{ihe2026, serranodata2026}. We begin by testing the assistance-free explanations with quantities that require no model fitting. These are the correlation between students' midterm and final scores, the spread of the within-student score changes, and a permutation test that asks whether the pairing of midterms with finals carries any information at all. We then fit our central model, a two-component mixture, which treats the class as a blend of two kinds of score pairs. In a coupled pair, both scores reflect the same student's proficiency, as ordinary measurement implies. In a decoupled pair, the two scores are statistically independent. The model estimates from the data which blend of the two kinds describes the class best. Production of exam answers by an outside source such as generative AI is one mechanism that generates decoupled pairs, though not the only one. Finally, we validate this model on simulated classes where the generating mechanism is known.

Three findings emerge. First, students' midterm and final scores show no detectable relationship to each other. The correlation between a student's midterm and final scores is 0.06. Under our models of the assistance-free explanations, it should be roughly 0.4 to 0.7. Score changes from midterm to final also spread about twice as widely as those models predict. Permutation tests detect no association between students' midterm and final scores. Simulated classes built from the assistance-free explanations essentially never look like the released data as long as both exams still track the same students' proficiency. The simulated classes begin to resemble the released data only when nearly all of that link is severed. Midterm answers that were not the students' own work, as with unauthorized AI use, would sever it. So would a final testing largely different material, with no assistance involved. The released scores cannot say which happened, and the mechanisms we examine are not exhaustive.

Second, nothing in the data requires the two exams to be related at all. A model in which a student's midterm carries no information about that student's final describes the released scores as well as any other candidate we examined. A closely matched model allowing weak dependence fits nearly as well, though not well enough to justify its extra parameter. Models that treat all 59 pairs as a single kind, describing each exam with one bell curve, fit far worse. A single bell curve cannot match midterm scores that pile up against the maximum, so this poor fit says as much about distributional shape as about whether the exams are related.

Third, our analysis shows what the evidence cannot do. The mixture model estimates the blend of coupled and decoupled pairs that best describes the class, but that blend is not a calibrated share of students, so we make no claim about how many students used assistance. Nor can the scores identify which students, if any, used it, or by themselves justify sanctions against anyone.

Beyond analyzing the Brown episode, we intend our work as a framework for future analyses. Generative AI is accessible to most students in higher education, and disputes that pair suspicious aggregate score patterns with contested alternative explanations and with institutional demands for individualized proof are likely to recur. The tools required to evaluate such disputes are not exotic: a scatterplot, a correlation, a permutation test, simulated benchmark classes, a mixture model checked against non-mixture alternatives, and a stress test on simulated classes where the truth is known. Section~\ref{sec:template} assembles these steps into a protocol that a future analysis can follow.

The rest of the paper is organized as follows. Section~\ref{sec:case} provides necessary background. Section~\ref{sec:related} reviews the literature that informs the assistance-free mechanisms. Section~\ref{sec:model} develops the measurement model, expresses the assistance-free explanations and the assistance mechanism as competing statistical accounts, describes estimation and testing, and collects every assumption in one place. Section~\ref{sec:results} presents results, and Section~\ref{sec:discussion} offers discussion and conclusions.

A note on how to read this paper: The methods are necessarily technical in places, but we have aimed to keep them accessible to a reader with a basic background in probability and statistics, introducing each statistical tool in plain language where it first appears. The paper also supports a lighter reading. Each results subsection opens by stating the question it answers and the answer, every figure and table caption does the same, and Section~\ref{sec:conclusions} gathers the conclusions in plain terms, so a reader can follow the argument without working through the machinery.

\section{Background and data}\label{sec:case}

Our analysis rests on the factual record assembled below. Throughout, we distinguish what the released data show directly from what is known only through press-reported aggregates. Section~\ref{sec:episode} recounts the episode as documented in the professor's own published account and contemporaneous reporting. Section~\ref{sec:data} describes the released student-level dataset, establishes its provenance, verifies it against independently reported statistics, and states the scope of our inputs.

\subsection{The incident}\label{sec:episode}

We rely on the following account, assembled from the professor's own first-person essay \citep{serrano2026fp}, an interview he gave to the Chronicle of Higher Education \citep{goldstein2026}, and contemporaneous reporting \citep{ihe2026, globe2026, chandonnet2026, bdh2026, gioino2026, tomshardware2026}. In the spring 2026 semester, following a December 2025 shooting on campus, Serrano offered the midterm examination of his advanced undergraduate course in welfare economics as a take-home for the first time. The exam was closed-book on the honor system, with an eleven-hour window for a test designed to take two hours \citep{serrano2026fp}. Enrollment, typically around 30, reached 86 after the format was announced \citep{ihe2026}. The midterm, administered March 5, produced a mean of 96, a median of approximately 98, and about 40 perfect papers \citep{gioino2026, bdh2026}. The professor reports designing it to be harder than the historical midterms, whose means ranged from 65 to 80 \citep{bdh2026, serrano2026fp}. Grading revealed stylistic anomalies. Many papers proved one result through an identical, convoluted argument by contradiction where a short direct argument was standard, and the same argument emerged when the professor submitted his questions to ChatGPT \citep{ihe2026, globe2026}. Study partners submitted essentially identical papers with identical errors \citep{bdh2026}.

The professor then announced that the final would be in-person, closed-book, and three hours long. He also pre-committed publicly to a decision rule under which the midterm would stand if the final's grade distribution roughly matched it, and would otherwise be voided, with the final reweighted to 80 percent of the course grade \citep{serrano2026fp}. Eighteen students dropped the course and nine more skipped the final; we refer to these 27 students as \emph{leavers}. Twenty-two of the leavers had scored a perfect 100 on the midterm \citep{goldstein2026, globe2026}. The professor voided the midterm, lowered the pass threshold to 40, and went public after describing the university's response to his complaint as inadequate \citep{goldstein2026, ihe2026}. Table~\ref{tab:reported} collects the press-reported aggregates, which provide context for the class-level analysis. The student-level data described next, covering the 59 students who completed the course (hereafter, \emph{finishers}), supersede them wherever the two overlap.

\begin{table}[htbp]

\centering
\footnotesize
\begin{tabular*}{\textwidth}{@{\extracolsep{\fill}}lcl@{}}
\toprule
Reported quantity & Value & Source \\
\midrule
Enrollment (typical) & 86 (${\sim}30$) & \citet{ihe2026} \\
Take-home window & 11 h for a 2-h exam & \citet{serrano2026fp} \\
Midterm mean, all 86 & 96 & \citet{serrano2026fp} \\
Midterm median & ${\sim}98$ & \citet{bdh2026} \\
Perfect midterm papers, all 86 & ${\sim}40$ & \citet{serrano2026fp} \\
Historical midterm means & 65--80 (easier exams) & \citet{serrano2026fp, bdh2026} \\
Dropped course / skipped final & 18 / 9 & \citet{ihe2026} \\
Perfect midterms among the 27 leavers & ${\sim}22$ & \citet{goldstein2026, globe2026} \\
Historical final means & minimum 65 & \citet{ihe2026} \\
\bottomrule
\end{tabular*}
\caption{Press-reported aggregates used for class-level context. Quantities describing the 59 finishers are computed directly from the released data (Table~\ref{tab:verify}) and are not repeated here. The reported counts 40 and 22 cannot both be exact, since the data contain 23 perfect midterms among finishers; we treat them as approximate.}\label{tab:reported}
\end{table}

\subsection{The released dataset}\label{sec:data}

The dataset analyzed in this paper contains the midterm and final scores of the 59 students who completed the course, anonymized as S1--S59. It is the data underlying the interactive chart embedded in Inside Higher Ed's July 8, 2026 article, which reports it as the data Serrano submitted to Brown's Standing Committee on the Academic Code in May 2026 \citep{ihe2026, serranodata2026}. We downloaded the dataset from that chart on July 16, 2026. No other student-level information is present. The 27 students who left before the final appear in no dataset; everything we say about them derives from the press-reported aggregates in Table~\ref{tab:reported}.

Table~\ref{tab:verify} checks the dataset against the separately published record. The agreement is close to exact. The data reproduce the reported 19 finishers below the pass threshold, the three zeros on the final, and the single student who improved. The reported ``only two within 10 points'' holds under a strict reading, since a third student's scores differ by exactly 10.0. Two small discrepancies remain. The final mean in the data is 48.8 against the widely reported 48.6. The finishers include 23 perfect midterms where two reported round numbers jointly imply 18, so at least one of those round numbers is approximate. Neither discrepancy affects our analysis, which uses the student-level pairs directly rather than any reported aggregate. We take this agreement as evidence that the file was extracted accurately and is consistent with the published record. Because the published accounts likely share a common origin, the agreement does not independently verify what the professor submitted, and no independent audit of the underlying gradebook exists.

\begin{table}[htbp]

\centering
\small
\begin{tabular*}{\textwidth}{@{\extracolsep{\fill}}lccl@{}}
\toprule
Statistic & In the data & Press report & Verdict \\
\midrule
Final mean & 48.8 & 48.6 & close; differs by 0.2 \\
Finals below 40 & 19 & 19 & exact \\
Finals of zero & 3 & ${\sim}3$ & exact \\
Students who improved & 1 & 1 & exact \\
Within 10 points of own midterm & 2 strict / 3 inclusive & 2 & consistent under strict $<10$ \\
Perfect midterms among finishers & 23 & 18 implied & press round numbers \\
\bottomrule
\end{tabular*}
\caption{Consistency of the released dataset with separately published statistics. ``18 implied'' arises from subtracting the reported 22 perfect-midterm leavers from the reported 40 perfect papers; since the data show 23 among finishers, at least one of the reported round numbers is approximate.}\label{tab:verify}
\end{table}

The released data and the press-reported aggregates of Table~\ref{tab:reported} are the only episode-specific sources for this study. The literature reviewed in Section~\ref{sec:related} enters the analysis only by informing how large we assume the assistance-free effects to be. We have had no contact with the professor, the university, or any student, and the analysis rests entirely on the public record. All conclusions are conditional on the accuracy of that record. Section~\ref{sec:assumptions} and the limitations discussion return to this dependence.

\section{Related work}\label{sec:related}

One line of academic-integrity research asks what a flag from an AI-text
detector does and does not establish.
\citet{weberwulff2023testing} tested fourteen detectors and found overall
accuracy below 80 percent, with roughly a fifth of AI-generated passages
misclassified as human-written and about half escaping detection once the
text was edited or machine-paraphrased. \citet{vanvlasselaer2026who} compared four
detectors on a constructed benchmark of 160 documents of known origin, with
human-written, AI-generated, hybrid, and AI-humanized papers in equal parts.
The result was real improvement: false positives were rare, and one tool
performed well even on the disguised texts. They nonetheless concluded that such tools
``should not be used as sole evidence in high-stakes decision-making.'' The
errors also fall unevenly across students. \citet{liang2023gpt} ran seven
detectors on 91 TOEFL essays by non-native English speakers and on 88 essays
by native-speaking American eighth graders: the detectors flagged more than
half of the TOEFL essays as AI-generated, a 61.3 percent average
false-positive rate, while scoring the eighth graders' essays nearly
perfectly. A flag can therefore reflect who a student is rather than what
the student did. \citet{eaton2023postplagiarism} argues that hybrid
human--AI writing will become the norm, making the attempt to determine where
the human ends and the machine begins pointless. Our evidence is score
patterns rather than text, but the same question confronts it: what, exactly,
does the signal establish?

The question of what evidence should be required before an institution acts
on suspicion is older than generative AI \citep{bretag2016handbook}. Writing
about contract cheating under the title ``(Im)possible to Prove,''
\citet{ellis2022impossible} show how academic judgment and bibliographic
forensics can supply evidentiary support in investigations, with procedures
built to improve probative value, fairness, and transparency. \citet{rogerson2017detecting} reached through casework the
conclusion we reach by computation: irregular patterns in student work raise
a question that the patterns alone cannot answer. The formal process itself carries
costs whatever its outcome: students who faced contract-cheating proceedings
describe the experience as ``mess, stress and trauma'' \citep{pitt2021mess}.
A large survey adds a caution about what aggregate signals may encode: among
14,086 Australian students, contract cheating was associated with speaking a
language other than English at home \citep{bretag2019contract}. The survey
establishes the association and nothing more. The hazard we take from it,
which the survey itself did not test, is that any procedure acting on
aggregate patterns can end up treating a correlated characteristic as though
it were conduct.

Whether assistance-free explanations can account for the Brown episode
depends on magnitudes: how many points an unproctored take-home format adds,
how much test anxiety subtracts, how much harder the final could have been.
Section~\ref{sec:assistance-free} gives each mechanism a number, and two
strands of literature bear directly on how large those numbers can plausibly
be.

The first strand of literature compares proctored with unproctored assessment. In a quasi-experiment spanning 648 students in two online courses whose other features remained essentially unchanged from prior terms, mean exam scores under webcam proctoring were 10 to 20 percentage points lower than in the earlier unproctored cohorts \citep{dendir2020}. The average differences were 18.6 percentage points in an economics course and 13.5 in a geography course. The authors attribute the difference to cheating that the camera deterred, though a cohort comparison cannot conclusively isolate that mechanism, and they note that some of the proctored deficit could reflect the added stress of being recorded. Because the cohort difference can combine cheating, proctoring stress, and other between-term changes, it is not an estimate of an assistance-free format effect, and we use it only as a deliberately generous benchmark. Meta-analytic evidence from ability testing points the same direction at smaller magnitude: across 49 studies and more than 100{,}000 test takers, unproctored assessments scored a pooled 0.20 standard deviations higher than proctored ones, with the advantage concentrated in tasks whose answers are easy to look up online \citep{steger2020meta}. In the small subset of those studies that tested the same people in both settings, the pooled correlation between a person's proctored and unproctored scores was 0.58, a level the authors read as considerable rank-order change, and far larger than the 0.06 in the released data.

A pandemic-era study reached essentially the same numerical association but interpreted it more favorably. Across 18 courses and roughly 2{,}000 students at a large midwestern American university whose assessment moved online, unproctored exam scores correlated with the same students' proctored in-person performance at 0.59, which the authors took to suggest that cheating was ``either not widespread or ineffective at boosting scores'' \citep{chan2023}. A published critique disputes that optimistic conclusion, pointing out that the study's own supplementary analyses show unproctored scores exceeding proctored ones by 0.4 standard deviations overall and 0.7 on multiple-choice questions \citep{newton2023}. In a systematic review of self-report surveys, 29.9 percent of students admitted cheating in online exams before the pandemic and 54.7 percent during it, figures the reviewers caution rest on surveys that likely underreport \citep{newton2024common}. For our purposes, the important observation is narrower. In that large sample predating public access to modern AI assistants, unproctored scores retained a moderate positive association with the same students' proctored performance---far from complete stability, but nothing like independence.

Mass outsourcing of assessment predates modern AI assistants. Student requests to the ``homework help'' site Chegg in five STEM subjects rose 196 percent between April--August 2019 and the same months of 2020, the first pandemic exam seasons \citep{lancaster2021}. Broad public access to modern general-purpose large language model assistants came later still, between the studies above and the episode analyzed here. Whether the association survives unproctored assessment in their presence is therefore an open question, and one the released data can answer for this one course. Princeton, for one, is not waiting for the answer: scheduled take-home final examinations there fell from 168 to 49 between spring 2025 and spring 2026, a decline of 71 percent \citep{princetonian2026}.

The second strand of literature concerns the psychological and format effects underlying the assistance-free explanations. A systematic review of take-home examination research finds them widely recommended for assessing higher-order skills such as analysis and evaluation, with the risk of unauthorized assistance their most cited disadvantage \citep{bengtsson2019}. The review is qualitative and reports direct comparative evidence as sparse, so it supplies no numerical estimate of an assistance-free take-home advantage; the twenty-point bonus we adopt in Section~\ref{sec:assistance-free} is a deliberately generous scenario value, not an estimate from this literature. The association between test anxiety and examination performance is modest. Across a synthesis of 562 studies, mean correlations are $-0.29$ with aptitude and achievement tests (grades 4 through postsecondary) and $-0.15$ with course grades \citep{hembree1988}. An update of that synthesis thirty years later puts the postsecondary achievement-test correlation at $-0.24$, characterizing effects across domains as ``small to moderate'' \citep{vonderembse2018}.

Our statistical approach also descends from an older literature, the
psychometric detection of testing irregularities \citep{cizek2017handbook}. Screening
multiple-choice exams for suspicious answer similarity,
\citet{wesolowsky2000detecting} modeled \emph{non}-cheating rather than
cheating behavior, because the innocent process requires the fewest
assumptions. He calibrated the method by simulation, and he named the
prevention of false accusation as the main concern. We make the same three
choices on a different kind of data, for the same reasons. The same
literature also speaks to how much methods of this kind can resolve. In the
simulations of \citet{wollack1997nominal}, an item-level copying statistic
built on an item-response model attained good power only once at least 20
percent of items were copied on an 80-item test, or 30 percent on a 40-item
test. Under a known response model, \citet{vanderlinden2006detecting} derived
the exact null distribution of chance answer matches so that the power of
the resulting test could be computed rather than guessed at. Those tests see every item a student answered; we see two
totals per student. Such item-level results suggest, though they do not
establish, that a pair of totals resolves less.

\section{Models and methods}\label{sec:model}

This section builds the analysis's machinery in the order its construction requires; Section~\ref{sec:results} will then present the resulting evidence in a different order, from least model-dependent to most. Section~\ref{sec:problem} explains why comparing the two exams' averages cannot settle the controversy and why the joint distribution of score pairs can. Section~\ref{sec:latent} sets out a standard measurement model for exam scores, and Section~\ref{sec:assistance-free} expresses each proposed assistance-free mechanism as an explicit parameter within it, anchored in the literature of Section~\ref{sec:related} where possible. Section~\ref{sec:mixture} introduces the mixture model that quantifies ability decoupling, Section~\ref{sec:estimation} describes estimation and uncertainty, Section~\ref{sec:modelcomp} sets out the model-comparison design, and Section~\ref{sec:modelfree} presents the permutation test and direct diagnostics, none of which uses the mixture. Section~\ref{sec:validation} stress-tests the method on simulated classes generated outside its own assumptions. For convenience, Section~\ref{sec:assumptions} summarizes all of our assumptions, and Table~\ref{tab:notation} collects the notation.

Across this section and the results, the analysis is organized around four questions. First, does the pairing of midterms with finals carry any information at all? The permutation tests of Section~\ref{sec:modelfree} answer this using no model. Strong association would mean the assistance-free accounts survive on this axis. Second, can the quantified assistance-free explanations reproduce the joint pattern of the released scores, even when we tune them to match both observed class averages exactly? The simulated benchmarks and the mean-matched design of Section~\ref{sec:modelfree} answer this. Third, does anything in the data require students' midterm and final scores to be related at all? The model comparison of Section~\ref{sec:modelcomp} answers this, with candidates designed so that no model wins merely by describing each exam's own score distribution more flexibly. Fourth, how should we summarize how far the scores depart from the shared-proficiency pattern of ordinary measurement, and what may that summary never be read as? The mixture model of Section~\ref{sec:mixture} provides the summary, and the simulation study of Section~\ref{sec:validation} establishes its limits.

\subsection{The inferential problem}\label{sec:problem}

Between the midterm and the final, at least five things changed at once. First, proctoring was added: students took the midterm unsupervised at home and sat the final in a supervised room. Second, the allotted time was reduced from eleven hours to three. Third, the professor raised the weight of the final from a fraction of the course grade to 80 percent of it. Fourth, the psychological context changed, from a take-home offered partly as a trauma accommodation to an in-person exam in a post-shooting semester. Fifth, the final may have been more difficult. Any comparison of the two averages is therefore confounded. But the released data contain far more than two averages. They contain the joint distribution of the two scores, that is, which student got what on both exams, and the assistance-free explanations, once quantified as in Section~\ref{sec:assistance-free}, make sharp predictions about that joint structure that averages alone do not determine.

\subsection{A measurement model for exam scores}\label{sec:latent}

Our foundation is the central postulate of classical test theory, arguably the oldest idea in educational measurement \citep{hambleton1993}: a student's score on any one exam is an imperfect reading of an underlying command of the material. We represent student $i$'s command of the material by a single number $\theta_i$ (``latent ability''), expressed directly on the 0--100 grading scale so it can be read as ``the score this student would earn on a typical proctored exam, averaged over many sittings.'' Near the scale's ends this reading is approximate, because the clamp introduced below pulls recorded averages away from $\theta_i$; that is exactly why boundary scores receive special treatment. We use ``ability'' in a deliberately narrow sense: throughout, $\theta_i$ denotes student $i$'s command of this course's material during the weeks spanning these two exams, and nothing more. It is a subject-specific, time-specific summary of prepared performance, not a fixed trait and not a measure of a student's overall intellect. Students learn, improve, and differ along many dimensions that no single number describes. The model requires only that this narrow quantity change little over the few weeks between the two exams, an assumption whose limits Section~\ref{sec:assistance-free} discusses.

Classical test theory turns this postulate into an equation by writing each score as the true score plus an error. For a properly proctored exam, where the recorded score reflects the student's own work, our version is
\begin{equation}\label{eq:proctored}
S_i \;=\; \mathrm{clamp}\!\left(\theta_i + \varepsilon_i\right),
\qquad \varepsilon_i \sim \mathcal{N}(0, \tau^2),
\end{equation}
where $\mathcal{N}(m, s^2)$ denotes a normal distribution with mean $m$ and standard deviation $s$, and $\varepsilon_i$ is \emph{occasion noise}: the day-to-day variation---sleep, luck in which topics appear, arithmetic slips---that makes the same student score differently on different sittings. We take the occasion noises to be independent across students and across exams. The additive true-score-plus-noise form is the standard classical-test-theory decomposition. The normal noise, with a single spread $\tau$ shared by all students, is our simplifying choice.

The clamp, $\mathrm{clamp}(x) = \min\{100, \max\{0, x\}\}$, is also a modeling choice, needed because nothing confines the sum $\theta_i + \varepsilon_i$ to the grading scale. The sum can exceed 100 for a strong student on a lucky sitting, or fall below zero for a struggling student on an unlucky one, and in either case the recorded score stops at the scale's end. We therefore read a recorded 100 as ``at least 100'' and a recorded 0 as ``at most 0.'' Treating scores pinned at the ends of a bounded scale this way, as censored observations, is standard: \citet{tobin1958} introduced the approach, and \citet{mcbee2010} recommends it specifically for test scores with floor or ceiling effects. Section~\ref{sec:mixture} makes this treatment precise.

Equation~\eqref{eq:proctored} describes each exam on its own. The controversy, though, turns on how the two exams relate, and here we make another key assumption: the midterm and the final are readings of the same $\theta_i$, so that in measurement terms the two exams act as parallel instruments for one construct, apart from a uniform shift in level that Section~\ref{sec:assistance-free} introduces. This assumption is what makes ``both exams measure the same student'' precise. Classical test theory itself regards true scores as test-dependent \citep{hambleton1993}, so nothing guarantees the assumption holds for these two exams. Both were written by the same instructor for the same course's material, but no item-level evidence of comparability is available. Section~\ref{sec:modelcomp} therefore includes among the fitted alternatives a general model that imposes no shared construct at all.

The model has two key parameters: the ability spread $\sigma$, the standard deviation of $\theta_i$ across the class, and the occasion noise $\tau$. Ability across the class, like the occasion noise, is modeled as normally distributed. Neither parameter can be estimated for this course from outside the model: no historical score distributions or test-retest records are available, and estimating both from the paired scores themselves would assume the shared-construct structure that is here in question, an exercise the free-noise fit of Section~\ref{sec:modelfree} performs as a check. We therefore assume values that produce realistic classroom behavior and then vary them to see how strongly the results depend on the choice. Before the clamp, a score in equation~\eqref{eq:proctored} is the sum of two independent parts, so the standard deviation of pre-clamp scores across a class is $\sqrt{\sigma^2+\tau^2}$. Setting $\sigma = 12$ and $\tau = 7$ makes that spread about 14 points. Much larger spreads would pile a sizable share of scores against the 100-point ceiling even in ordinary years. This reasoning is a plausibility check rather than an estimate---reported means alone do not determine spreads---so the sensitivity analysis repeats everything across a range of assumed values.

The same ability-plus-noise split determines how strongly two exams that both read $\theta_i$ should correlate across students. The two exams share each student's ability but draw independent occasion noises. Ability therefore contributes all of the shared variance, $\sigma^2$ out of a total of $\sigma^2 + \tau^2$, and the pre-clamp correlation between the two exams equals that share, $\sigma^2/(\sigma^2+\tau^2) \approx 0.75$. In measurement terms this ratio is the exams' implied \emph{reliability}, a quantity we return to repeatedly. At the assumed values, durable differences among students account for three quarters of the variance in scores, and day-to-day factors account for the rest. In recorded scores, the clamp shrinks both the spread and the correlation below these pre-clamp values. Every simulation and fitted model below includes the clamp, so no result relies on the pre-clamp formulas.

To assess how strongly the results depend on our baseline choices of $\sigma$ and $\tau$, we re-fit the central model of Section~\ref{sec:mixture} over the full grid of assumed values, $\sigma \in \{9, 12, 15\}$ crossed with ten noise values $\tau \in \{4, 6, 7, 8, 10, 12, 14, 16, 18, 20\}$, and we sweep the simulated assistance-free benchmark of Section~\ref{sec:modelfree} over the same ten values (Section~\ref{sec:mixfit}). Equation~\eqref{eq:proctored} also supplies an intuitive check on our baseline values. In an ordinary year, with nothing pushing all scores up or down between the two sittings, a student's two scores differ only through the two occasion noises, and that difference has standard deviation $\sqrt{2}\,\tau \approx 9.9$ points. The model then implies that about 69 percent of students would land less than ten points from their own midterm on the final. In the data, 2 of 59 do.

Equation~\eqref{eq:proctored} already contains two statistical artifacts that featured prominently in public discussion of the episode. The first is the \emph{ceiling effect}, produced by the clamp. When many students score near 100, the exam stops distinguishing among them, scores pile up at the top, and the observed variance shrinks. The second is \emph{regression to the mean}. Occasion noise is drawn independently on each exam, so a student who scored unusually well the first time, partly through good luck, will on average score closer to their own $\theta_i$ the second time. Extreme measurements tend to be followed by less extreme ones even when nothing about the person has changed. The clamp is built into everything that follows, and regression to the mean into every account that retains shared ability. Every simulation and every fitted model carries the clamp, with boundary scores treated as the censored observations described above, and regression to the mean is part of the joint structure that the simulations and the central model share (Section~\ref{sec:mixture}). Both artifacts work in the assistance-free explanations' favor: the ceiling compresses high scores, and regression to the mean pulls a lucky high midterm back down on the final, with no assistance involved in either. Building both in keeps the analysis fair to those explanations. Any drop or decoupling that the two artifacts can generate on their own appears in the assistance-free simulations and fits, so the artifacts, as modeled, are not mistaken for evidence of assistance. One qualification applies. Regression to the mean requires a stable $\theta_i$ for a student's scores to return to, so among the alternative models of Section~\ref{sec:modelcomp}, those that relax the shared-ability structure lose that mechanism, while the shared-shift candidates keep it.

\subsection{Assistance-free mechanisms as model parameters}\label{sec:assistance-free}

We turn each assistance-free explanation into an explicit, adjustable model parameter. Some magnitudes can be anchored in the literature of Section~\ref{sec:related}, and there we use the published values. For the rest we assume a \emph{scenario value}, chosen deliberately large so that the explanation gets every benefit of the doubt. Once an explanation carries a number, it predicts specific score patterns that the data can check. Each parameter enters as a mean shift that moves every student's expected score by the same amount. Treating the effect as the same for every student is itself a modeling choice, not a consequence of writing the explanation as a parameter. A 20-point bonus, for example, raises every student's pre-clamp score distribution by 20 points; the recorded class average rises by less when papers press against the 100-point ceiling, and occasion noise varies the realized scores. The mean-matched design of Section~\ref{sec:modelfree}, which calibrates the location and total shift to the observed averages, does not depend on this distinction. Changes that instead differ from student to student are a separate scenario, which the simulation study of Section~\ref{sec:validation} simulates directly. Three parameters cover the proposed mechanisms. We take up a fourth mechanism, motivation, at the end of this subsection.

The first parameter captures the possibility that a student honestly scores higher on a take-home exam because of the format itself: more time, less pressure, the chance to check one's work. Take-home midterm scores therefore receive a bonus $B$, so that student $i$'s midterm score is $M_i = \mathrm{clamp}(\theta_i + B + \varepsilon_{i1})$. The largest published estimate of the unproctored advantage is 10 to 20 points, and that estimate includes the cheating that proctoring deters \citep{dendir2020}. A pure format effect should therefore be smaller, which makes $B = 20$ a generous extreme.

The second and third parameters concern the final. This final may have been more difficult than the historical finals, and sitting a proctored three-hour exam in a post-shooting semester carries a psychological cost. An extra-difficulty parameter $D$ lowers every final score equally, and an anxiety-and-format parameter $A$ does the same, so that student $i$'s final score is $F_i = \mathrm{clamp}(\theta_i - D - A + \varepsilon_{i2})$. Neither parameter has a clean literature anchor. The anxiety studies of Section~\ref{sec:related} measure how more-anxious students differ from less-anxious ones, not what happens to scores when a whole class's testing conditions change, so we adopt $A = 6$ as a scenario value. The extra difficulty $D$ has no external anchor at all, so we adopt $D = 12$ as another. Little rides on the exact scenario values. Large shifts push scores against the floor and ceiling of the grading scale, where the clamp distorts spreads and correlations, so no conclusion below rests on an assumed shift size. The strongest form of the analysis concedes the assistance-free account whatever location and shift reproduce the two observed class averages (Section~\ref{sec:modelfree}), and the mixture model estimates the combined shift freely in any case.

The fourth mechanism is student motivation, and we deliberately leave it out. One might propose that students tried less hard on the final, but the incentives ran the other way. The professor had announced in advance that the final could become 80 percent of the course grade, which raised the stakes well above those of an ordinary year. Nor could a student coast on a banked midterm score, because the pre-announced decision rule put that score at risk. Under that incentive reading, omitting a motivation penalty is conservative: if higher stakes raised effort, final scores under the assistance-free account should if anything have been better than we model. High stakes do not mechanically raise performance---some students may choke or disengage---but motivation effects that differ from student to student are a form of heterogeneous individual change, which the simulation study of Section~\ref{sec:validation} examines directly. The three blank finals, which may reflect surrender rather than effort, are the exception. The limitations discussion returns to them.

The key structural fact is that each mechanism above, in the form proposed in public discussion and modeled here, shifts all students together. Format, difficulty, and anxiety apply the same expected change $-(B + D + A)$ to every score, while class composition and attrition determine who is in the data but not how any individual's two scores relate. Selecting who remains can still move class-level statistics such as the observed correlation, and the simulated benchmarks include exactly that effect (Section~\ref{sec:modelfree}). We call the resulting simulated benchmark the \emph{homogeneous shared-shift benchmark}. In it, a plot of final against midterm scores forms a cloud whose long axis runs along a line of slope one: every student drops by roughly the same amount, give or take the occasion noise and the pull of regression to the mean. That prediction requires no estimation, and Section~\ref{sec:modelfree} tests it directly against the released scores. The benchmark commits to one version of each public explanation: a shift of a stated size, applied to every student alike. Other versions are conceivable, and the simulation study of Section~\ref{sec:validation} probes departures from the benchmark directly.

A natural objection is that proficiency itself can change between the two exams, as when some students study harder and improve while others disengage and decline. With only two exams, modest individual change of this kind cannot be told apart from occasion noise, so the noise term, which permits within-student swings on the order of ten points, absorbs it statistically, even though genuine learning or decline is not noise. What the shared-shift structure excludes is proficiency change that is both very large and different from student to student. Rather than dismissing that possibility, we test it: the simulation study of Section~\ref{sec:validation} generates classes with no decoupled pairs whose students' proficiencies genuinely change by different amounts, with standard deviations up to 20 points, and asks whether such classes reproduce the released pattern. A structural observation says what to watch for in those results: individual changes add noise to each student's scores but largely preserve each student's relative standing, so breaking the rank order requires changes that are either very large relative to the spread of ability or tied to a student's prior standing, and the study includes generators of both kinds.

\subsection{A mixture model for ability-decoupled scores}\label{sec:mixture}

To summarize how far the class's joint score distribution departs from the coupled structure of ordinary measurement, we use a finite mixture model, a standard statistical tool for describing a population composed of unobserved subgroups, called \emph{latent classes} in mixture terminology \citep{mclachlan2019}. The model treats the 59 finisher score pairs as a blend of two kinds. Rather than classifying any individual pair, it asks what blend of the two kinds best explains all 59 at once. In a \emph{coupled} pair, both exams measure the same student, so the pair inherits the joint structure that equation~\eqref{eq:proctored} implies. Before clamping, the two scores are jointly normal,
\begin{equation}\label{eq:coupled}
\begin{pmatrix} M_i^* \\ F_i^* \end{pmatrix} \sim
\mathcal{N}\!\left(\begin{pmatrix} m_u \\ m_u - S_u \end{pmatrix},\;
(\sigma^2+\tau^2)\begin{pmatrix} 1 & \rho \\ \rho & 1 \end{pmatrix}\right),
\qquad \rho = \frac{\sigma^2}{\sigma^2+\tau^2},
\end{equation}
and recording clamps each coordinate to the grading scale. Here $m_u$ is the coupled component's midterm mean, and $S_u$ is its combined uniform shift, the sum $B + D + A$ of Section~\ref{sec:assistance-free}: the take-home bonus is lost, and the difficulty and anxiety penalties are gained. The data can identify the three shifts only in this combination, and the subscript $u$ is mnemonic for \emph{uniform shift}.

Equation~\eqref{eq:coupled} carries the full structure of Section~\ref{sec:latent}. The correlation $\rho$ arises because both exams share the student's $\theta_i$. Regression to the mean is built in: the conditional mean $\mathbb{E}[F_i^* \mid M_i^*] = m_u - S_u + \rho\,(M_i^* - m_u)$ pulls a student whose midterm ran high partly on luck back toward the center. The within-student change $M_i^* - F_i^*$ has mean $S_u$ and standard deviation $\sqrt{2}\tau$, because the shared $\theta_i$ cancels while the two independent occasion noises both contribute.

In a \emph{decoupled} pair, the model treats the midterm as carrying no information about the same student's final---the pattern expected when, for example, exam answers are produced by an outside source---so the component makes the two scores statistically independent:
\begin{equation}\label{eq:decoupled}
M_i^* \sim \mathcal{N}(m_M, s_M^2),
\qquad
F_i^* \sim \mathcal{N}(m_F, s_F^2),
\end{equation}
again clamped to the grading scale, with all four parameters estimated from the data rather than assumed. Section~\ref{sec:latent} promised that boundary scores would be treated as censored observations, and this is where that treatment takes effect: when a model is scored against the data, a recorded 100 counts as ``at or above 100'' rather than as an exact reading, and likewise at 0. A pair recorded at both boundaries, such as a perfect midterm alongside a blank final, contributes the joint probability of both boundary events. This is how the model represents the pile-up of perfect papers at the ceiling.

The model's final ingredient is the mixture weight $q$. A class built from the blend draws each score pair from the decoupled component with probability $q$ and from the coupled component with probability $1-q$, so $q$ describes the recipe for the class as a whole. At the endpoints its meaning is clean: $q = 0$ says the class follows the coupled account alone, and $q = 1$ the independent component alone. Between the endpoints, a fitted value is a \emph{descriptive index}: it locates the class's joint score distribution between the two fitted templates. That is all we claim for it, and three misreadings deserve explicit warning. The weight says nothing about individual students, because the model never records which component produced which pair and classifies no one. Under the model's own assumptions, $q$ is the probability that a pair comes from the decoupled component; even so, a fitted weight does not reliably estimate the true fraction of decoupled pairs, because in simulated classes where that fraction is known by construction, the fitted weight need not track it (Section~\ref{sec:valresults}). And it is not a dial that measures how much association remains between the two exams. Even a blend whose scores are independent within each component can produce association across the class, because the two components sit in different parts of the score scale: a pair's midterm then hints at which component the pair came from, and through that hint carries information about its final. The model comparison of Section~\ref{sec:modelcomp} includes a variant designed to isolate exactly this effect. One built-in asymmetry deserves its own flag. The coupled component inherits the fixed spreads and correlation of Section~\ref{sec:latent}, while the decoupled component's four marginal parameters are fitted freely, so a class whose two margins differ sharply---as ceiling-heavy midterms beside widely spread finals do---can push weight toward the decoupled component partly through marginal fit alone. This is precisely the confound the comparison of Section~\ref{sec:modelcomp} is built to separate, and one more reason the weight is a descriptive index. With these warnings in place, the fitted weight is the summary the rest of the paper tracks; the diagnostics of Section~\ref{sec:modelfree} assess the same pairing question without using the mixture at all, so no conclusion rests on the weight alone.

Mixture formulations of this kind, in which a latent class separates aberrant from ordinary test behavior, are established tools in psychometrics, where they detect cheating on compromised items and item preknowledge \citep{shu2013, xi2025}. Our model applies the same logic at the level of whole exam papers rather than individual items.

Two interpretive points matter. First, the component labels describe statistical structure, not conduct. A decoupled pair is one whose midterm carries no information about the same student's proctored performance under the model. That pattern says nothing about its cause. Wholesale outsourcing to generative AI can decouple scores, but so can old-fashioned collusion, and the professor's account describes some of the latter \citep{bdh2026}. Other causes are logically possible as well; Section~\ref{sec:modelcomp} therefore compares alternative statistical structures, with no one-to-one correspondence between candidates and causes. Nor is the fitted weight causally identified: the episode offers no randomization or instrument that could support causal claims, and we make none. It is a model-based description of the released score pairs. Second, the mixture contains the quantified shared-shift account of Section~\ref{sec:assistance-free} as the special case $q = 0$. The coupled component carries the same uniform shift $S_u$ introduced in Section~\ref{sec:assistance-free}, fitted freely rather than fixed in advance. If the data needed no decoupled component, the fit would return $q$ near zero.

\subsection{Estimation and uncertainty quantification}\label{sec:estimation}

We estimate the mixture by maximum likelihood: among all settings of the model's seven fitted parameters, we search for the one under which the observed 59 pairs would be least surprising. The weight $q$ is the summary the paper tracks throughout. The other six describe the two components' locations and spreads, and Table~\ref{tab:notation}, at the end of this section, collects the full notation and marks which quantities are fitted and which are assumed. The rest of this subsection quantifies how sure we can be about the fitted weight, in three ways: a bootstrap analysis of its stability, a profile of the likelihood, and checks of its dependence on individual observations. Section~\ref{sec:modelcomp} then compares the mixture against alternative models.

Each pair's contribution to the overall likelihood is a weighted blend of how probable that pair is under the two components, $q \cdot (\text{decoupled}) + (1-q) \cdot (\text{coupled})$, with boundary scores handled as the censored observations of Section~\ref{sec:mixture}. Finding the best-fitting settings is a numerical search over a landscape of fit quality, and for mixtures that landscape can have several peaks: a search can stall on a foothill (a local optimum) rather than the summit. We therefore maximize the log likelihood by the Nelder--Mead simplex method \citep{neldermead1965} from 40 random starting points and keep the best result.

The search also needs a guard. Maximum likelihood rewards a model for making the observed data probable, and an unconstrained normal mixture can exploit that reward in a degenerate way. If one component's mean is centered on a single observed score and that component's spread shrinks toward zero, the model becomes arbitrarily sure of that one observation, and the likelihood grows without limit. The resulting fit is perfect on paper and describes no real class. We therefore require each fitted spread, a standard deviation, to lie between 2 and 50 points. The lower bound rules out the collapse just described, and the upper bound merely keeps the search inside a range wide enough for any plausible score distribution.

To quantify uncertainty in the estimate we use the bootstrap \citep{efron1979}. The question it answers is how sensitive the fitted weight is to the particular score pairs in the data: a sturdy estimate should not hinge on a handful of pairs. Only the one class was observed, so we approximate that sensitivity by resampling. We rebuild the class 2{,}000 times, each time drawing 59 whole score pairs with replacement from the observed 59, so a student's pair may appear twice in one rebuilt class and not at all in another. Resampling whole pairs keeps each student's two scores together. We refit the model to each rebuilt class and take the central 95 percent of the refitted $q$ values, the interval from their 2.5th to their 97.5th percentile, as a stability range. The 59 finishers are not a probability sample from any defined population of classes, so the range describes stability under perturbation of this class, not inference to a larger population.

Refitting 2{,}000 rebuilt classes invites the same local-optimum problem as the original fit. Each resample is first refit starting from the full-data solution and from three fresh random starts. Because such refits can still miss the best fit, especially near the $q = 1$ boundary, every replicate is then re-optimized from ten additional widely dispersed starting points, and an independent audit repeats the search from ten more starts at every hundredth replicate to gauge how often the protocol stops short. For each replicate, the corrected record keeps the best fit found at any stage. We release the initial refits alongside the corrected record, and Section~\ref{sec:mixfit} reports how much the corrections change the range. Two caveats explain why we report a stability range rather than a confidence interval. The estimate sits near the top of its allowed range (a weight cannot exceed 1), and near such a boundary the usual reading of such an interval as a confidence interval, one that would capture the truth about 95 percent of the time, is unreliable. The range also takes the component forms and assumed constants as given, so it reflects resampling variability rather than model uncertainty.

A second view of the same uncertainty comes from profiling the likelihood. We refit the model with $q$ held fixed at each value on a grid and record where the fit deteriorates, which shows which values of the weight the data can and cannot accommodate within this mixture. We treat the profile as exploratory. The usual recipe for turning a profile into a confidence interval relies on textbook conditions that fail here: at $q = 1$ the coupled component describes no one and its parameters lose meaning, so the profile supports no formal coverage claim. Whether the observed pairing of midterms with finals shows more association than random pairings do is tested nonparametrically instead, by the permutation test of Section~\ref{sec:modelfree}, which involves no model and no optimizer.

Last, we ask whether the estimate is a fact about the class or an accident of a few students. With 59 pairs, one unusual pair could in principle move the fitted weight by itself, and a class-level conclusion should not hinge on any single student. We therefore refit the model 59 times, deleting one whole score pair at a time, always pairs rather than single scores. We also refit with the few pairs whose two scores lie less than ten points apart removed, because those pairs look most like ordinary coupled measurement and this refit shows how much the fit leans on them. We report only aggregate summaries of these refits. A per-student influence value would say how much one pair moves a class-level fit, not whether that student did anything, so such values could not identify misconduct any more than the rest of the analysis can. We decline to produce them anyway, because a ranked table of per-student numbers would invite exactly that misreading.

\subsection{Comparing candidate models}\label{sec:modelcomp}

We compare the mixture against eight alternative models. A model of paired scores makes claims about two different things at once. The first is each exam's \emph{margin}: the shape of that exam's own score distribution viewed alone, as if the scatterplot of pairs were projected onto a single axis---for the midterm, a tall pile of perfect papers with the rest spread below. The second is the \emph{dependence}: how the two scores pair up within students. A model earns likelihood credit on both fronts at once, and that creates a trap: a candidate can beat a simple independence benchmark not because its account of the pairing is better, but merely because its margins describe each exam's own histogram better. The comparison is therefore built like a controlled experiment: candidates vary in whether the exams are coupled and in how flexible their margins are, with the key candidates matched on marginal flexibility, so that remaining differences in fit can be traced to what each model says about the pairing. In plain terms, the comparison asks: once a model is allowed to describe each exam's own score distribution well, does anything in the data \emph{require} a student's midterm score to be related to that same student's final score?

Each of our eight candidate models is designed to answer a specific question. The first three keep the exams coupled. The calibrated shared-shift model is our quantified version of the assistance-free account: one population, everyone shifting together, the two exams coupled as strongly as the measurement model of Section~\ref{sec:latent} implies, with the spreads and correlation fixed at the assumed constants and only the location and the common shift fitted. It asks whether the account fits at its full prescribed coupling strength. The flexible shared-shift model frees the overall spread and the between-exam correlation, asking whether the account fits once the data choose the strength of the relation. The general censored bivariate normal, with all five of its parameters free (two means, two spreads, and a correlation), asks whether \emph{any} jointly normal one-population account fits: any linear relation, each exam described by a single bell curve.

Three other candidate models make the two exams statistically independent, asking whether independence fits once each exam's own distribution is described well. The simplest describes each exam by one censored normal. A second lets the midterm margin be a two-component censored-normal mixture, matching the marginal flexibility of the main model at the same parameter count. The third, the most closely matched comparator, is the product of the main mixture's two marginal distributions, which we call the product model: the same marginal families as the main model, with no dependence of any kind. Multiplying the two margins is the mathematical form of statistical independence, because every question about a pair is then answered by consulting each exam's own distribution separately. The product model is the comparison's key benchmark: it keeps the mixture's own margins while discarding everything the mixture says about the pairing, so if the mixture fits no better than the product of its own margins, then whatever advantage the mixture holds over other candidates came from its margins, not from any information in the pairing.

The product model needs one further precaution in fitting. Its final margin is itself a small mixture, and censoring creates an escape route: the search can always improve the fit a little more by sliding a tiny, low-weight component of final scores ever further below the score floor, chasing the handful of finals recorded at zero. No best-fitting setting exists for the search to converge to, and the drifting fit reflects an artifact of censoring rather than any real class. We therefore fit the model twice: once unconstrained, reported as a diagnostic and excluded from our ranking of model fits, and once with its component means constrained to the score scale. The constrained fit anchors all comparisons, and we call it the constrained product model. Constraining the means this way is a regularization choice rather than neutral bookkeeping, and the constrained optimum still presses one component mean against the boundary, so the anchor is a boundary-adjacent fit rather than an interior one. Section~\ref{sec:modelcompresults} reports how the anchor's ranking moves when the constraint moves and explains why we draw no conclusion from fine orderings among the leading candidates.

Two further candidates sit between the coupled and independent groups. The first is a variant of the mixture in which the correlation inside the coupled component is forced to zero, so that within each of the two components, a student's midterm carries no information about the same student's final. Even so, the class as a whole retains the subtle form of association that Section~\ref{sec:mixture} described: because the two components occupy different parts of the score scale, a pair's midterm hints at which component the pair came from, and that hint shifts expectations for the final. We therefore call this candidate the within-class independence variant rather than an independence model, and it exists to measure how much of the mixture's fit flows through that membership channel alone.

The second candidate addresses a gap in the set so far: between the full shared-ability coupling of the measurement model and no dependence at all, nothing yet offers middle ground. This candidate lets the data choose a \emph{small} amount of dependence. It starts from the constrained product model, keeping the same descriptions of each exam's own scores, and adds one new dial using a Gaussian copula, a standard statistical device for adding a chosen strength of association between two quantities while leaving each quantity's own distribution untouched. The dial runs from strongly negative association through zero to strongly positive, and it captures association of the familiar kind, in which higher midterms go with consistently higher, or consistently lower, finals. (Technically, the dial is a correlation on an underlying normal scale rather than the Pearson correlation of the recorded scores.) Set to zero, this candidate is exactly the constrained product model; freed, the dial and all the marginal parameters are re-estimated together. The comparison is therefore a clean nested one: whatever fit the candidate gains over the product model measures exactly what one dial's worth of dependence is worth. Two fitting details carry over from the product model: scores at 0 or 100 are treated as censored, exactly as in every other candidate, and the fitted copula model, like the constrained product model, ends with the mean of one small component pressed against the bottom of the score scale---the boundary behavior described above, which matters when Section~\ref{sec:modelcompresults} compares these two fits finely.

A model with more parameters fits at least as well within a nested family, and more flexible candidates generally gain fit from their extra freedom, so raw likelihoods cannot be compared directly. We report the Akaike information criterion \citep[AIC;][]{akaike1974}, which balances fit against the number of parameters, with its small-sample correction AICc. Lower values indicate a better tradeoff. We read AICc differences strictly as relative support within this candidate set, not as hypothesis tests and not as identification of a mechanism. We also read the differences loosely, for a reason already encountered: AICc's penalty counts each free parameter as one full unit of flexibility, but several leading fits end with a parameter pushed to the edge of its allowed range, and a parameter stuck at an edge does not use its full unit, so the count is only an approximation of how flexible each model really is; weakly identified, nearly empty mixture components strain the approximation further. As a rough conventional guide, candidates within about two AICc units of the leader are effectively tied, and candidates more than about ten units behind have essentially no support \citep{burnham2004}. Section~\ref{sec:modelcompresults} reads the observed differences against that scale with these cautions in force. The comparison's endpoint is deliberately modest: it ranks fit--complexity tradeoffs among these candidates, and it cannot show that the top candidate is true, that the candidate set is exhaustive, or that any particular behavior produced the scores.

\subsection{Diagnostics that do not use the mixture model}\label{sec:modelfree}

None of the diagnostics in this subsection uses the mixture. They lean on assumptions to three different degrees, and we present them in that order: a permutation test that uses no model at all, simulated benchmarks that use the measurement model but fit nothing to the observed pairs, and a final check that fits the shared-shift account directly.

The permutation test asks the most assumption-light question available for paired data: does the pairing matter at all? To answer it, we shuffle the final scores against the midterms, which destroys any within-student relationship while preserving both score distributions exactly, and we ask where the observed statistics fall among the shuffled ones. No model, assumed constant, or optimizer is involved. We use three statistics. The Pearson correlation and the Spearman rank correlation detect linear and monotone association. The distance covariance \citep{szekely2007} is zero only under independence, so its permutation test is sensitive to dependence of any form. If both exams measure the same students and retain substantial shared signal relative to occasion noise, as the public assistance-free explanations combined with the measurement model imply, the observed statistics should sit far outside their permutation distributions. One qualification travels with every resampling method in this paper. The permutation argument is exact under the null hypothesis of no association when the 59 rows are exchangeable, that is, when no row is systematically tied to another; the reported $p$-values approximate that argument with large numbers of random shuffles. The public record describes near-identical collaborative submissions, which would violate exchangeability, so we read the $p$-values as approximate, and the same caution applies to the case-resampling bootstrap of Section~\ref{sec:estimation}.

One further check accompanies the permutation test, because failing to reject independence is not the same as showing independence: a real link between the exams could simply be too weak for 59 pairs to reveal. To measure what the test can and cannot detect, we simulate classes in which the true strength of the between-exam link is known and dialed from zero to strong: each simulated pair of scores shares a \emph{latent correlation}, set before the ceiling clamps scores, that we sweep from 0 to 0.7. The clamp and attrition then make the recorded correlation smaller than the latent one, just as they do in the released data. Each simulated class resembles the released one in every other respect: its averages are calibrated to match both observed class averages, scores pile against the 100-point ceiling through the same censoring, and 27 of 86 students leave by the same attrition rule. Running the two-sided Pearson permutation test on 1{,}000 classes at each dial setting, and recording how often it detects the link at the 5 percent level, traces the test's \emph{power curve}, reported in Section~\ref{sec:mfresults}. Because this check relies on simulated classes, it shares the assumptions of the second level, described next, rather than the permutation test's freedom from them.

The second level uses the measurement model as a benchmark. The uniform-shift structure of Section~\ref{sec:assistance-free} yields two diagnostics that we compute directly from the data, with nothing fitted: the Pearson correlation $r(M, F)$ between the two exams, and the standard deviation of the within-student changes $M_i - F_i$. The predictions are simple. If every student shifts by roughly the same amount, students who scored high on the midterm should also score high on the final, so $r(M, F)$ should be large. And each student's change should equal the shared shift plus the difference of two occasion noises, so the spread of changes should be near $\sqrt{2}\,\tau$, about 9.9 points at the baseline values. Once censoring and attrition act, the simulated recorded spread is nearer 11 points, and all benchmarks below account for this. The benchmarks themselves come from the model: we simulate classes from the assistance-free account and record both quantities in each simulated class. These are therefore pre-specified simulation checks rather than assumption-free tests, and Section~\ref{sec:assumptions} lists the assumptions they use.

To avoid resting the comparison on any single configuration, we sweep the simulated benchmark over a grid. The grid varies the occasion noise $\tau$ over the ten values of Section~\ref{sec:latent}, the ability spread $\sigma$ over $\{9, 12, 15\}$, and the class ability center $\mu$ (the mean of latent ability $\theta_i$) over $\{65, 72.5, 80\}$, holding the shifts at their generous values $B = 20$, $D = 12$, $A = 6$. The grid also crosses two attrition mechanisms. Each starts a simulated class at 86 students and removes 27 to leave the observed 59: one removes the 27 students of highest latent ability, and the other removes the 27 highest \emph{observed} midterms. The second mechanism matches the reported fact that most leavers held perfect midterms, and because it selects on noisy scores rather than on ability, it generates additional regression to the mean, which favors the assistance-free account. In every cell of the grid---each combination of noise, spread, center, and attrition mechanism---we record how often a simulated class is at least as extreme as the released data on \emph{both} diagnostics at once; this joint criterion prevents the account from passing each test at a different noise level. The criterion is one-sided by design. A simulated correlation at or below the observed value counts, including strongly negative ones, because the benchmark predicts positive association and the question is how rarely it falls to the observed level or beyond. The simulation study of Section~\ref{sec:validation} additionally records a symmetric variant, which asks the different question of whether a simulated class \emph{resembles} the observed near-zero association.

The grid above holds the shifts at their scenario values, and we call it the \emph{raw grid} for that reason. Because of the clamp, a different location or a much larger shift could in principle alter the diagnostics, so we add a second, stronger design that dispenses with the scenario values altogether. For every noise level, ability spread, and attrition mechanism, we calibrate the latent midterm location and the total shift so that the simulated classes reproduce the observed midterm mean and the observed final mean. The calibration is performed once per cell, by simulation: a grid search chooses the location and shift so that the average simulated class, after censoring and attrition, reproduces both observed means, and those values are then fixed for all 2{,}000 classes in the cell, with the achieved simulated means reported in Table~\ref{tab:matched}. This concedes the assistance-free account both observed averages and a shift with no literature-derived bound, and only then do we apply the same joint diagnostic. The calibration matches the two averages rather than the full score distributions, so the diagnostic tests the pairing structure together with the distributional shape the calibration leaves free. We record the achieved distributional features for every cell, so the residual mismatch stays visible. All tail frequencies are counts out of 2{,}000 simulated classes per grid cell. The third level is a final check that fits rather than simulates: we fit the shared-shift model directly to the score pairs, using the censored likelihood, with the location, the shift, and the noise $\tau$ all free, so that the account may claim as much noise as it needs. We report the $\tau$ it requires.

\subsection{Stress tests on simulated classes and dependence sensitivities}\label{sec:validation}

We measure the analysis's behavior on simulated classes where the truth is known. One distinction matters throughout: ``no unauthorized assistance'' is not the same as ``no statistical decoupling.'' Most of the designs below generate every score from the simulated student's own abilities and preserve substantial signal between the two exams, so that both exams still carry shared information about the same students. For those designs, the simulated classes serve as negative controls. If the released configuration rarely arises from such classes, then mechanisms of that kind, at the magnitudes and in the forms we simulate, rarely produce it. One design instead weakens the statistical link itself: the construct-overlap design lets the two exams measure diverging constructs, and at overlap zero the final is statistically \emph{independent} of the midterm by construction. Those cells contain full statistical decoupling produced by an assistance-free mechanism, and we include them precisely to show that the statistical pattern does not identify its cause.

The primary question we ask of each simulated class involves no model fitting at all, and it comes in two variants. The one-sided variant of Section~\ref{sec:modelfree} asks whether the class's correlation is at or below the observed value \emph{and} its change spread at least as large as observed. The symmetric variant asks whether the class resembles the observed near-zero association: an absolute correlation no larger than the observed 0.061, with the same spread condition. Because these joint criteria require no optimizer, their answers cannot be artifacts of numerical fitting. Secondarily, we fit the mixture to every simulated class and report the \emph{full distribution} of the fitted index, not merely how often it crosses a threshold.

We use four designs. The first adds heterogeneous genuine change: each student's shift from midterm to final is drawn with its own value, $S_i \sim \mathcal{N}(\bar S, \omega^2)$, with the heterogeneity $\omega$ swept from 0 to 20 points, so that students may truly improve or decline by very different amounts.\footnote{Before attrition, the $\omega = 0$ generator coincides with the mixture's own $q = 0$ boundary. The attrition step then selects rows by an order-statistic rule, so even this cell reaches the fitted family only approximately, and every other generator lies outside it.} The second lets the two exams measure diverging constructs: each student receives a second ability for the final, jointly normal with the first and correlated $\lambda$ with it, with $\lambda$ swept from 0.9 down to 0, so that at $\lambda = 0$ the two abilities, and hence the two exams, share nothing. So that this design is not handicapped by construction, we calibrate the second ability's spread to make simulated final scores match the observed final-score spread. The last two designs violate the independence of rows. One duplicates students in duos whose two rows share everything, ability, occasion noises, and shift alike, producing identical rows that roughly halve the amount of independent information in a class while leaving its apparent size unchanged. The other gives duos of distinct students a single shared midterm built from the average of both partners' abilities, rescaled to preserve the ability spread, while each student's final reflects that student's own ability. This collaboration-like design is the one design whose scores are not generated purely from each student's own traits, and we include it as a row-dependence sensitivity, with its authorization deliberately left unspecified, rather than as an assistance-free account. Every cell uses the baseline noise, $\tau = 7$, applies censoring and observed-midterm attrition, and is calibrated once per cell, as in Section~\ref{sec:modelfree}, to reproduce both observed average scores; complete generator definitions are in the released pipeline.

We simulate 300 classes per cell and fit each from twenty-four starting points spanning the full range of $q$, including informed low- and mid-$q$ starts built from each simulated class's own means and spreads. We release every replicate's fitted index and diagnostics. Finally, because a numerical search can stop short of the best fit, we double-check every fit by re-running it from thirty additional widely dispersed starting points; whenever a re-run finds a better fit, the better fit replaces the original, and all summaries are computed from these corrected fits.

\subsection{Summary of assumptions}\label{sec:assumptions}

Each assumption below was introduced and justified where it arises in Sections~\ref{sec:data} and \ref{sec:latent}--\ref{sec:validation}; we gather them here so that they can be seen at once. They are not equally important. A1 and A2 concern the inputs. A3 and A8 mix structural modeling choices with assumed values probed by sensitivity analysis. A4 and A5 are the structural assumptions a critic should examine first, A6 and A7 set the scope of what we claim, and A9 underlies every fitted likelihood and the resampling $p$-values. The evidentiary layers use different subsets of the list. The permutation test of Section~\ref{sec:modelfree} uses only A1 and A9, and the simulation benchmarks there use A1 through A4 with A7 and A8. Only the fitted models use the rest, so a reader unpersuaded by the mixture machinery can rely on the earlier layers alone. Each entry states where it is probed, and a reader can replace any assumption and rerun the analysis using the released code.

\begin{enumerate}
\item[\textbf{A1.}] \emph{The released dataset is accurate}, and the press-reported aggregates in Table~\ref{tab:reported} are approximately accurate. Every conclusion is conditional on this. Probed by: the consistency checks in Table~\ref{tab:verify}; press aggregates are treated as approximate throughout.
\item[\textbf{A2.}] \emph{Boundary scores are censored readings.} Recorded scores are mechanically bounded to $[0, 100]$; treating each recorded 0 or 100 as a censored reading of a latent value at or beyond the boundary is a modeling assumption shared by every fitted candidate, not itself a mechanical fact. The permutation tests of Section~\ref{sec:modelfree} do not use it.
\item[\textbf{A3.}] \emph{Occasion noise is independent across students and exams}, normal, with $\tau = 7$. A class-wide shock on final day is \emph{not} assumed away---it is absorbed by the free shift $S_u$. Probed by: the full assumed-constant sweep of the mixture over $\tau = 4$--$20$ (Section~\ref{sec:mixfit}); the null sweep of $\tau$; the free-$\tau$ fit of Section~\ref{sec:modelfree}; and the heterogeneous-shift and clustered-submissions simulation cells (Section~\ref{sec:validation}). Not probed: the normal form of the noise, and the assumed zero correlation between a student's two occasion errors, which if negative would widen score changes while weakening association.
\item[\textbf{A4.}] \emph{Coupled pairs follow the parallel-forms, shared-shift structure} of equation~\eqref{eq:coupled}, with normal ability spread $\sigma = 12$. Treating the two exams as parallel measures of one construct, and each student's $\theta_i$ as stable across the weeks between them, are substantive assumptions (Section~\ref{sec:latent}). Probed by: the general bivariate alternative in the model comparison, which imposes neither parallel forms nor a shared shift, and the heterogeneous-shift and construct-overlap simulation cells (Section~\ref{sec:validation}), which quantify what happens when this assumption fails.
\item[\textbf{A5.}] \emph{Decoupled pairs are independent censored normals}, equation~\eqref{eq:decoupled}, with all parameters fitted. The defining assumption is independence, that is, decoupling. The censored-normal margins are fitted forms, and nothing is assumed about the source or quality of any outside output. Probed by: a sensitivity fit that frees the within-component correlation (Section~\ref{sec:mixfit}).
\item[\textbf{A6.}] \emph{The 59 finishers are informative only about finishers.} Class-level statements would require both assumptions about the 27 unobserved leavers and a literal reading of the mixture's latent classes, which the data do not identify; we therefore make no quantitative class-level claims (Section~\ref{sec:classlevel}).
\item[\textbf{A7.}] \emph{No motivation penalty on the final}, a choice that is conservative under the incentive reading of Section~\ref{sec:assistance-free}: the final's stakes rose to 80 percent of the course grade, and heterogeneous responses to those stakes are probed separately. We retain the three blank finals in the main fit and exclude them in sensitivity.
\item[\textbf{A8.}] \emph{Assistance-free parameter values} for the simulated null. $B = 20$ is anchored at the largest published unproctored advantage, which itself includes deterred cheating. $D = 12$ and $A = 6$ are generous scenario values rather than literature-derived bounds, since the anxiety literature reports individual-difference correlations, not class-wide causal effects. The simulated nulls also assume that attrition removes the 27 leavers by one of two mechanisms, the highest latent abilities or the highest observed midterms; richer selection is not modeled. Probed by: the raw null grid, which sweeps $\tau$, $\sigma$, $\mu$, and the attrition mechanism, and the mean-matched null, which discards the scenario values entirely and calibrates the location and total shift to the two observed average scores.
\item[\textbf{A9.}] \emph{Rows are independent draws.} Every fitted likelihood multiplies across the 59 rows, treating each student's score pair as an independent draw. The resampling inferences use the related, weaker condition of exchangeability: the permutation $p$-values and the case-resampling bootstrap treat the 59 rows as interchangeable units, with no row systematically tied to another. The public record describes near-identical collaborative submissions, so we read the resulting $p$-values as approximate (Section~\ref{sec:modelfree}). Probed by: the duplicate-row and shared-product simulation cells (Section~\ref{sec:validation}).
\end{enumerate}

\begin{table}[htbp]

\centering
\footnotesize
\begin{tabular*}{\textwidth}{@{\extracolsep{\fill}}clcl@{}}
\toprule
Symbol & Meaning & Value & Basis \\
\midrule
$M_i, F_i$ & student $i$'s midterm, final score & data & \S\ref{sec:data} \\
$M_i^*, F_i^*$ & pre-clamp midterm, final score & --- & eq.~\eqref{eq:coupled} \\
$\theta_i$ & latent ability (0--100 scale) & --- & \S\ref{sec:latent} \\
$\sigma$ & ability spread across students & 12 (9, 15) & assumed value, \S\ref{sec:latent}; A4 \\
$\tau$ & occasion (day-to-day) noise per exam & 7 (4--20 sweep) & implied reliability 0.75; A3 \\
$\rho$ & between-exam correlation, coupled pairs & $\sigma^2/(\sigma^2{+}\tau^2)$ & eq.~\eqref{eq:coupled} \\
$B, D, A$ & take-home bonus; final difficulty; anxiety & 20; 12; 6 & \S\ref{sec:related}; A8 \\
$q$ & mixture weight, decoupled component (descriptive index) & fitted & \S\ref{sec:mixture} \\
$m_u, S_u$ & coupled midterm mean; combined uniform shift & fitted & eq.~\eqref{eq:coupled} \\
$m_M, s_M$ & decoupled midterm mean, spread & fitted & eq.~\eqref{eq:decoupled} \\
$m_F, s_F$ & decoupled final mean, spread & fitted & eq.~\eqref{eq:decoupled} \\
\bottomrule
\end{tabular*}
\caption{Notation. ``Fitted'' parameters are estimated by maximum likelihood from the 59 released score pairs; values in parentheses are sensitivity settings. The combined uniform shift sums the three mechanism parameters, $S_u = B + D + A$; the data identify only the sum.}\label{tab:notation}
\end{table}

\section{Results}\label{sec:results}

We present the core evidentiary results in order of increasing model dependence, and then turn to their limits and interpretation. Section~\ref{sec:glance} describes the released data. Section~\ref{sec:mfresults} reports the permutation tests and simulation diagnostics, which show that the pairing carries no detectable information and that the homogeneous shared-shift benchmark built from the public assistance-free explanations rarely reproduces the data's joint pattern except at very low exam reliability. Section~\ref{sec:mixfit} reports the mixture-model fit with its uncertainty and sensitivity analyses, and Section~\ref{sec:modelcompresults} the comparison against the candidate models. The remaining subsections interpret rather than escalate: Section~\ref{sec:classlevel} explains why the analysis does not extend to the full class of 86, Section~\ref{sec:valresults} reports the simulation study, and Section~\ref{sec:cannot} states what the evidence cannot support. So that the thread stays visible, each subsection below opens with the question it answers.

\subsection{Descriptive analysis}\label{sec:glance}

\begin{figure}[htbp]
\centering
\includegraphics[width=\textwidth]{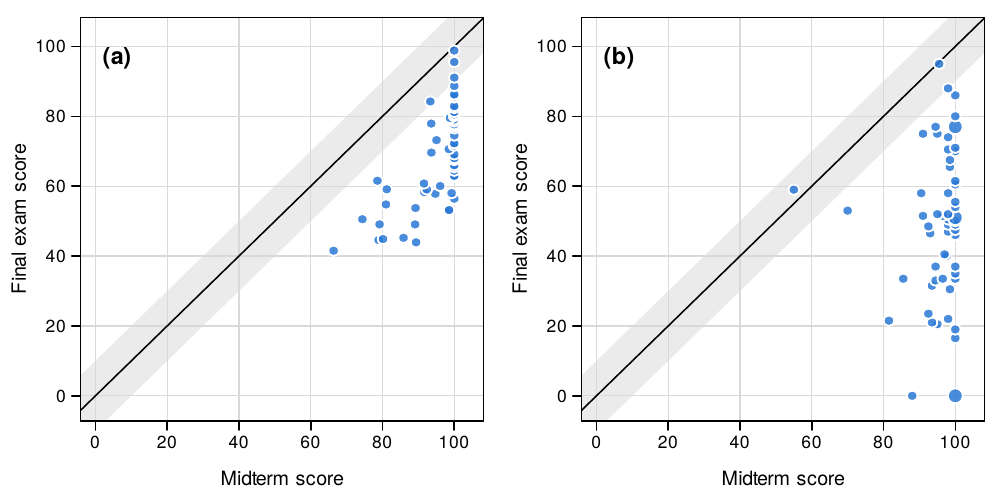}
\caption{What should the released class look like if the assistance-free explanations were right? Like panel (a): scores drop, but students keep their rank order. Panel (b) shows what the class actually looks like: scores drop and rank order all but disappears. \textbf{(a)} One simulated class from the shared-shift assistance-free model at generous settings (take-home advantage $B=20$, extra final difficulty $D=12$, anxiety $A=6$, occasion noise $\tau=7$, ability spread $\sigma=12$; average ability 86.5 chosen so the expected simulated midterm mean matches the observed value; the 27 highest observed midterms removed). In this particular draw, the midterm mean is 95.1, 32 midterms sit at the ceiling, and $r = 0.68$. \textbf{(b)} The released data: 23 midterms sit exactly at 100 while the same students' finals range from 0 to 86, and $r = 0.06$. The simulated class is matched to the observed midterm mean but not the final mean, so the panels differ in final-score level as well as in pairing; the mean-matched design of Section~\ref{sec:modelfree} makes the comparison that matches both averages. The diagonal line marks equal scores; the shaded band is $\pm 10$ points; marker area scales with the number of identical score pairs.}
\label{fig:scatter}
\end{figure}

\begin{figure}[htbp]
\centering
\includegraphics[width=0.85\textwidth]{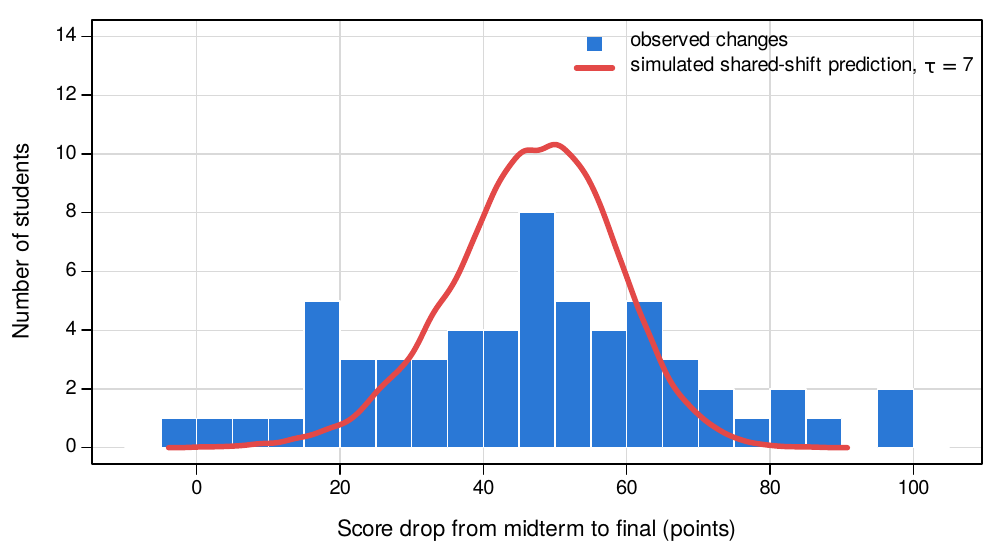}
\caption{How large should score changes be if every student shifted together? About half as spread out as the changes actually observed. Bars show the within-student changes across the 59 finishers, recorded as midterm minus final, so positive values are drops and the one negative value is an improvement. The curve shows what the uniform-shift assistance-free account predicts at the baseline occasion noise ($\tau = 7$, with the location and shift calibrated to the observed averages): the distribution of recorded changes estimated from 500 simulated classes, each simulated with censoring and observed-midterm attrition and matched to both observed average scores, scaled to the class size of 59. Under that account, recorded changes concentrate near a single shared shift, with standard deviation about 11.4. The observed changes have standard deviation 22.8, about twice the prediction, and range from $-4$ to $100$.}
\label{fig:gaps}
\end{figure}

The question here is descriptive: what do the released score pairs look like, side by side with what an assistance-free class should look like? Figure~\ref{fig:scatter} shows both. The simulated assistance-free class forms a correlated cloud, with high-midterm students landing high on the final and every student sitting a roughly constant distance below the diagonal. The released class does not. There, 54 of 59 midterms exceed 90 and 23 sit exactly at 100, while the same students' finals spread from 95 down to zero, and across the class the two scores are nearly unrelated. The Pearson correlation is $r = 0.061$, below the 5th percentile of the correlations the shared-shift assistance-free account produces and well inside what random pairing produces (Section~\ref{sec:mfresults}). The Spearman rank correlation is 0.12. Among the 36 students below the midterm ceiling, a subsample free of midterm censoring apart from one final at the floor, the correlation is 0.09, though restricting attention to a narrow midterm range can itself shrink a correlation. The within-student changes, recorded as midterm minus final so that positive values are drops, range from $-4$ (one student improved) to $100$ (a perfect midterm followed by a zero final), with mean 46.9 and standard deviation 22.8 (Figure~\ref{fig:gaps}). Two students' finals land less than ten points from their midterms, and a third differs by exactly ten.

\subsection{Simulation-diagnostic results}\label{sec:mfresults}

We now address the two questions: does the pairing of midterms with finals carry any information at all, and can the quantified assistance-free account reproduce the released joint pattern? The answers are that no test detects any information in the pairing, and that the account reproduces the pattern only when the exams are assumed to be mostly noise.

The permutation tests come first, because they use no model at all. Across 100{,}000 shuffles of the finals against the midterms, the observed correlation of 0.061 is entirely unremarkable: the two-sided $p$-value is 0.65 (0.36 for the Spearman version), and the central 95 percent of permuted correlations spans $[-0.25, 0.26]$: with 59 pairs, an observed Pearson correlation must fall outside that span before that test can register it. The omnibus distance-covariance test, sensitive to dependence of any form, gives $p = 0.62$ across 10{,}000 shuffles. By every statistic examined, then, the pairing of these two exams shows no detectable dependence. Nonrejection is not a demonstration of independence---with 59 pairs, weak dependence would escape these tests---but any account in which both exams substantially measure the same students must explain why no test registers it.

How much does that nonrejection actually rule out? The power curve of Section~\ref{sec:modelfree} answers with a direct experiment: we hand the test simulated classes whose true link strength we know, and we count how often it notices. Recall that the latent correlation is the link built into the score pairs before the ceiling clamps them, and that the recorded correlation comes out smaller. When the true link is zero, the test false-alarms in 4.6 percent of 1{,}000 classes, close to its nominal 5 percent level, so the ceiling pile-up and attrition do not break its calibration. When the true link is weak, the test usually misses it: at a latent correlation of 0.2, which the ceiling and attrition shrink to a recorded correlation near 0.16, the test detects the link only 23 percent of the time, and at 0.3, only 46 percent. But when the link is as strong as the modeled assistance-free accounts predict, the test almost always finds it: detection runs at 93 percent at a latent correlation of 0.5, where the recorded correlation is about 0.41, the bottom of the predicted range, and is nearly certain from 0.6 upward. The nonrejection above therefore leaves genuinely weak links open, and it counts as real evidence against a link of the predicted strength.

Under the shared-shift assistance-free model at the baseline values ($\tau = 7$, $\sigma = 12$), the simulated correlation between midterm and final among 59 finishers is $0.59$ on average, and 90 percent of simulated classes fall in $[0.43, 0.72]$. None of 2{,}000 simulated classes produces a correlation as low as the observed $0.061$, none produces a spread of score changes as large as the observed 22.8, and so none matches both. Doubling the occasion noise to $\tau = 14$ drags the simulated correlation down to $0.26$ on average, and about 6 percent of simulated classes then reach a correlation as low as observed; but the spread diagnostic still almost never matches (11 of 2{,}000 classes reach the observed spread), and only 6 of 2{,}000 match both at once. Sweeping the raw grid shows where the account can succeed: joint matches appear at appreciable rates only in its extreme-noise corner, at best 41 percent of classes at $\tau = 20$, $\sigma = 9$, $\mu = 72.5$. That configuration implies a test--retest reliability of 0.17, meaning that day-to-day noise, not durable differences among students, would account for 83 percent of score variance. Its simulated midterms, with mean 84 and 28 percent perfect papers, look nothing like the released ones, with mean 95.7 and 39 percent perfect.

The mean-matched design addresses what the raw grid leaves open. Conceptually, it hands the shared-shift account exactly the two class averages it must explain and then asks whether the pairing pattern follows. Table~\ref{tab:matched} reports it at the baseline spread, $\sigma = 12$. In every cell, we calibrate the location and total shift so that simulated classes reproduce the observed midterm mean of 95.7 and final mean of 48.8. The shift carries no literature-derived bound, and every calibrated value falls well inside the numerical search range. No cell with occasion noise up to $\tau = 14$ produces a single joint match in 2{,}000 classes. Matches first appear at $\tau = 16$ and grow as the noise rises. Across all sixty mean-matched cells, including the smaller and larger ability spreads, no configuration with implied reliability above 0.47 produces a single match. Matches become common only when the exams are mostly noise: the frequency reaches 1 percent only below implied reliability 0.41 and exceeds 10 percent only at 0.36 and below, exam scores that are roughly two parts noise to one part signal.

\begin{table}[htbp]

\centering
\footnotesize
\begin{tabular*}{\textwidth}{@{\extracolsep{\fill}}cccccccc@{}}
\toprule
 & Implied & \multicolumn{2}{c}{Achieved midterm} & \multicolumn{2}{c}{Achieved final} & \multicolumn{2}{c}{Joint matches per 2{,}000} \\
$\tau$ & reliability & mean & perfect & mean & zero & leavers by ability & leavers by midterm \\
\midrule
4 & 0.90 & 95.7 & 49\% & 48.8 & 0.0\% & 0 & 0 \\
6 & 0.80 & 95.7 & 51\% & 48.8 & 0.0\% & 0 & 0 \\
7 & 0.75 & 95.6 & 52\% & 48.6 & 0.0\% & 0 & 0 \\
8 & 0.69 & 95.8 & 54\% & 49.1 & 0.0\% & 0 & 0 \\
10 & 0.59 & 95.6 & 56\% & 48.9 & 0.1\% & 0 & 0 \\
12 & 0.50 & 95.8 & 60\% & 48.7 & 0.2\% & 0 & 0 \\
14 & 0.42 & 95.7 & 61\% & 48.8 & 0.4\% & 0 & 0 \\
16 & 0.36 & 95.7 & 64\% & 48.7 & 0.8\% & 7 & 10 \\
18 & 0.31 & 95.7 & 66\% & 49.5 & 1.1\% & 80 & 65 \\
20 & 0.26 & 95.8 & 68\% & 48.9 & 1.9\% & 315 & 188 \\
\bottomrule
\end{tabular*}
\caption{Can the shared-shift account, granted both observed class averages, reproduce the released pairing pattern? Only when the exams are assumed to be mostly noise: matches first appear at $\tau = 16$ (implied reliability 0.36), and across all sixty cells, spanning $\sigma \in \{9, 12, 15\}$, the joint frequency reaches 1 percent only below implied reliability 0.41. Each row is a mean-matched simulation at the baseline ability spread ($\sigma = 12$): we calibrate the class's average ability and the total shift so that simulated classes reproduce the observed midterm mean (95.7) and final mean (48.8), and achieved means land within 0.12 and 0.68 points of those targets. The calibration matches means only. The observed data have 39 percent perfect midterms and 5.1 percent zero finals, while the simulated cells overproduce perfect midterms and underproduce zero finals, a mismatch the design leaves visible rather than adjusts; the columns shown remove leavers by observed midterm, and the remove-by-ability variant is similar. Joint-match entries count simulated classes, out of 2{,}000, at least as extreme as the released data on both diagnostics at once ($r \le 0.061$ and change standard deviation $\ge 22.8$), a one-sided criterion in the direction opposite to the positive association the benchmark predicts (Section~\ref{sec:modelfree}).}\label{tab:matched}
\end{table}

Three qualifications apply. These are frequencies on the examined grid, not on every conceivable configuration. The calibration matches the two averages rather than the full score distributions, so the diagnostic tests the pairing structure together with whatever distributional shape the calibration leaves free; a simulated class could in principle fail the spread criterion partly because its final margin is too narrow rather than because it retains too much signal. The simulation study of Section~\ref{sec:valresults} closes part of that gap: its construct-overlap cells calibrate the simulated final scores' spread to the observed spread, so their near-parallel cells cannot fail the spread criterion merely through a too-narrow final margin, and those cells still produce no matches (Table~\ref{tab:valid}). And the reliability in question is the model's variance ratio $\sigma^2/(\sigma^2+\tau^2)$ rather than an observed correlation, though evidence from courses predating generative AI anchors its plausible range: unproctored and proctored scores in the same course then correlated strongly \citep{chan2023}. Letting the location, shift, and noise all float freely in the full censored model gives the same picture: the best single-shift description of the score pairs requires $\tau \approx 17.3$, an implied reliability of $0.33$. To give that number concrete meaning: before any ceiling effects, a reliability of 0.33 would mean that two sittings of the same exam by the same class correlate only 0.33, with about two thirds of the pre-clamp score variation reflecting the day rather than durable differences among students; recorded scores, compressed by the clamp, correlate somewhat less still (Section~\ref{sec:latent}). The shared-shift assistance-free account thus reproduces the two diagnostics at appreciable rates only by making the exams mostly noise. The model comparison of Section~\ref{sec:modelcompresults}, which uses all of the data rather than these two summaries, asks whether that much noise describes the data better than a mixture does.

\subsection{Mixture model estimates}\label{sec:mixfit}

The diagnostics established that the pairing carries no detectable information. The question for this subsection is what the mixture makes of that: permitted any blend of coupled and decoupled structure, how heavily does the best description of the 59 pairs weight the decoupled component, and how stable is that weight? The answer: almost entirely, and stably at the baseline constants and across all moderate-noise settings, softening only at the low end of the resampling range and under assumed values that make the exams mostly noise. As always, we read that weight as the descriptive index of Section~\ref{sec:mixture}, never as a share of students.

Table~\ref{tab:fit} reports the maximum-likelihood fit. The fitted mixture assigns weight $\hat{q} = 0.965$ to its decoupled component: permitted any blend of the two templates, the model reproduces the released data best when nearly all of the mixture's weight rests on the decoupled one. We report this as a \emph{descriptive index} of the fitted blend---under this model, the best description of the class is almost entirely the independent component---not as an estimated share of students and not as a graded measure of remaining dependence (Section~\ref{sec:mixture}): the model comparison of Section~\ref{sec:modelcompresults} shows that the data do not require the mixture's latent pair classes or establish that they exist, and the simulation study of Section~\ref{sec:valresults} shows that the index is not calibrated as a prevalence estimator. Calibrated here carries its estimator sense: a calibrated index, computed on classes where the true decoupled share is 30 percent, would return values centering near 0.30, and Section~\ref{sec:valresults} shows this index does not behave that way.

\begin{table}[htbp]

\centering
\small
\begin{tabular*}{\textwidth}{@{\extracolsep{\fill}}llc@{}}
\toprule
Parameter & Interpretation & Estimate \\
\midrule
$q$ & decoupled-component weight (descriptive index) & 0.965 \\
 & \quad 95\% case-resampling stability range (not a confidence interval) & [0.64, 1.00] \\
 & \quad exploratory profile-likelihood range & [0.90, 0.99] \\
 & \quad assumed-constant sweep, $\tau \le 14$, all $\sigma$ & 0.956--0.966 \\
$S_u$ & coupled-component uniform shift (points) & 6.8 \\
$m_u$ & coupled-component midterm mean & 63.6 \\
$m_M$ ($s_M$) & decoupled midterm mean (spread), pre-censoring & 98.6 \; (6.0) \\
$m_F$ ($s_F$) & decoupled final mean (spread) & 48.1 \; (23.3) \\
\bottomrule
\end{tabular*}
\caption{How does the fitted mixture describe the released class, and how sure can we be? Almost entirely through its decoupled component, with real but bounded uncertainty. Maximum-likelihood mixture fit to the 59 released score pairs ($\sigma = 12$, $\tau = 7$; jointly censored bivariate likelihood). The weight $q$ is a descriptive index, not an estimated share of students: Table~\ref{tab:models} shows the data do not require the mixture's latent pair classes, and Section~\ref{sec:valresults} shows the index is not calibrated as a prevalence estimator. The stability range is the corrected 95\% case-resampling range (all 2{,}000 replicates re-optimized from ten additional wide starts), not a confidence interval, and the profile range is exploratory. With nearly all of the mixture's weight on the decoupled component, the coupled component's parameters draw on little effective information and are weakly identified. All ranges are conditional on the component forms; dependence on the assumed constants is shown in Figure~\ref{fig:calib}. The index is $0.963$ with blank finals excluded.}\label{tab:fit}
\end{table}

\begin{figure}[htbp]
\centering
\includegraphics[width=0.8\textwidth]{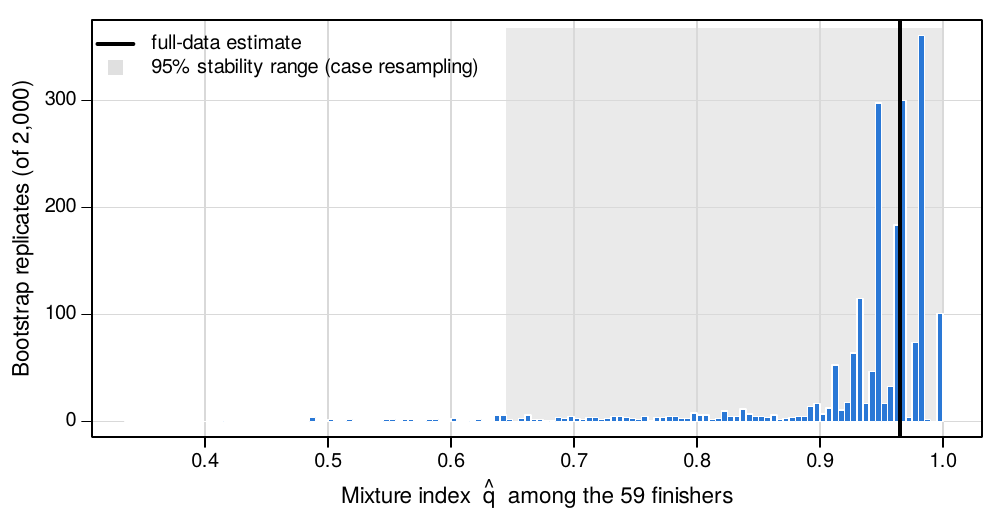}
\caption{How stable is the fitted index when the class is rebuilt from its own pairs? Fairly stable: most rebuilt classes keep nearly all weight on the decoupled component, though a minority land lower, making the range wide. Case-resampling distribution of the mixture index $\hat{q}$ among the 59 finishers (2{,}000 resamples of the released pairs, model refit to each). The bars show the corrected refits, with every replicate re-optimized from ten additional wide starting points. The vertical line marks the full-data index, 0.965; the shaded band is the corrected 95\% stability range $[0.64, 1.00]$, which is not a confidence interval with a guaranteed capture rate (Section~\ref{sec:mixfit}). Most rebuilt versions of the class still place nearly all of the mixture's weight on the decoupled component; a minority land lower, which is why the range is wide and why we do not read the index as precise.}
\label{fig:boot}
\end{figure}

Resampling the 59 pairs 2{,}000 times and refitting yields a case-resampling stability range, computed from the best fit found for each replicate. Because a numerical optimizer can stop short of the best fit, we re-optimized every one of the 2{,}000 refits from ten additional widely dispersed starting points, and an independent audit refit every hundredth replicate from ten more. The corrected record keeps the best fit found for each replicate. In that record, 257 replicates improve by more than 0.01 log-likelihood units, the threshold we treat as material (the largest improvement is 13.9 units), the pile of refits at $q > 0.999$ shrinks from 202 to 99, and the corrected 95 percent range is $[0.64, 1.00]$ (Figure~\ref{fig:boot}). Even after the main re-optimization, the audit's extra starts still improve one replicate, so optimizer noise persists at the single-replicate level; corrections of the observed sizes do not move the range. The width is itself informative: the data only weakly pin the index down, though boundary effects, leftover optimizer noise, and the coarseness of resampling 59 pairs all contribute to it. An exploratory likelihood profile concentrates far more tightly, over $[0.90, 0.99]$. That interval collects the weights at which the profiled log likelihood stays within 1.92 units of its best value, the cutoff a textbook 95 percent interval would use; the textbook guarantee itself does not hold here (Section~\ref{sec:estimation}). The two ranges answer different questions: the resampling range asks how the index moves when the class's composition is perturbed, while the profile asks how quickly the fit deteriorates for this class when the index is forced away from its best value. Near the boundary neither carries its textbook guarantee (Section~\ref{sec:estimation}), so we emphasize the qualitative reading: even at the low end of the resampling range, the fitted mixture still places about two thirds of its weight on the decoupled component.

\begin{figure}[htbp]
\centering
\includegraphics[width=0.85\textwidth]{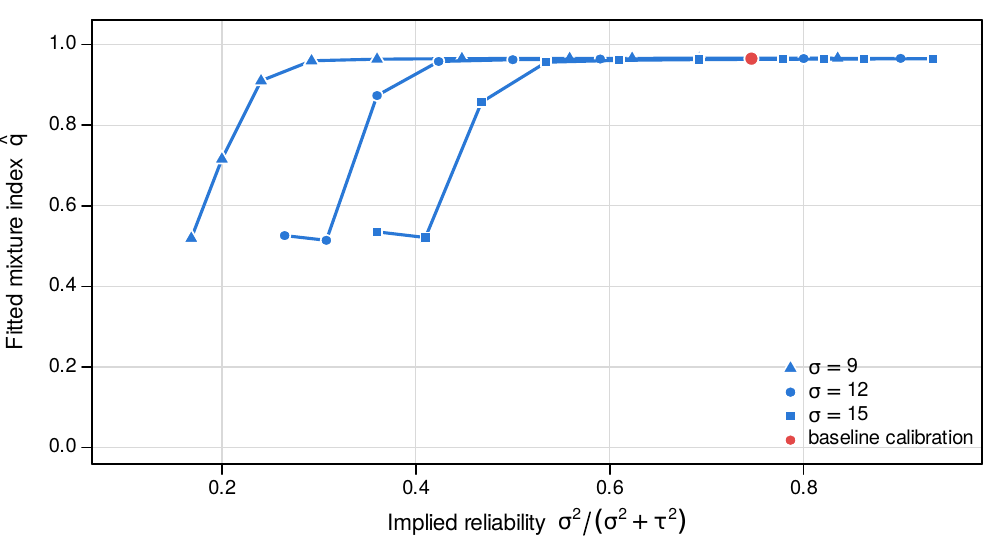}
\caption{How much does the fitted index depend on the assumed constants? Little, until the assumed values make exam scores mostly noise. Fitted mixture index $\hat{q}$ across the full grid of assumed constants, $\tau$ from 4 to 20 crossed with $\sigma \in \{9, 12, 15\}$, plotted against each cell's implied reliability $\sigma^2/(\sigma^2+\tau^2)$, the share of score variance due to durable student differences. The red marker is the baseline values, $\sigma = 12$ and $\tau = 7$. The index exceeds 0.95 in every cell with $\tau \le 14$ and falls toward one half only at values that make exam scores mostly noise, the same region in which the mean-matched assistance-free null of Table~\ref{tab:matched} first succeeds. The abrupt drops at low reliability mark assumed-value combinations where a second solution near one half overtakes the high-index fit (Section~\ref{sec:mixfit}).}
\label{fig:calib}
\end{figure}

The index's dependence on the assumed constants is itself a result, shown in Figure~\ref{fig:calib}. Recall that $\sigma$ is the assumed spread of durable ability differences across students and $\tau$ the day-to-day noise in any one exam sitting. Refitting the mixture over the complete grid, the ten $\tau$ values of Section~\ref{sec:latent} crossed with $\sigma \in \{9, 12, 15\}$, gives $\hat{q}$ between 0.956 and 0.966 in every cell with $\tau \le 14$. At $\tau = 16$ the index dips to 0.86--0.91, and at $\tau = 18$--$20$ fits near one half appear with slightly better log likelihood, as the likelihood surface flattens into a broad ridge. Two qualifications frame this pattern. Cells with equal implied reliability but different $(\sigma, \tau)$ can give different indices, so no single reliability threshold determines the answer, and we report the full grid. And the record from courses predating generative AI, while it documents strong correlation between unproctored and proctored scores \citep{chan2023}, supplies no sharp numerical bound on this model's variance ratio. What can be said is that the low-index fits require assumed values under which exam scores are roughly two-thirds noise. That is the same regime the mean-matched null of Section~\ref{sec:mfresults} needs and the free-noise fit selects, and a reader must judge whether course examinations behave that way.

The fitted coupled-component shift is $\hat{S}_u = 6.8$ points, well below the combined generous scenario values for format, difficulty, and anxiety ($B + D + A = 38$) and of a size those mechanisms could plausibly produce, though with $\hat{q} = 0.965$ this lightly weighted component rests on little information (Table~\ref{tab:fit}). The fitted decoupled component has midterms piled against the ceiling (mean 98.6 before censoring) and finals scattered wide (mean 48.1, standard deviation 23.3). The index is 0.963 when the three blank finals are dropped entirely, and it moves by less than $10^{-5}$ when each interior score is treated as known only to within its half-point rounding band. No single observation drives the index: deleting each pair in turn moves $\hat{q}$ by at most 0.018, and deleting the two pairs whose scores lie less than ten points apart raises the index to 0.979. A far simpler statistic tells the same story: 57 of 59 finishers' two scores differ by ten points or more, a share of 0.966. The ten-point threshold has no theoretical connection to the mixture, so its numerical agreement with $\hat{q} = 0.965$ is incidental; the point of the comparison is that the headline magnitude does not depend on the mixture machinery, and that neither number is a calibrated estimate of assistance prevalence. Assumption A5 is probed directly: work produced with partial outside input need not be fully decoupled---editing an outside product, for example, partially re-couples the two scores---so we refit the mixture with the decoupled component's within-pair correlation free rather than fixed at zero. The freed fit returns $\hat{q} = 0.964$ with a fitted correlation of 0.18, and the two specifications are not strongly distinguished, the freed version trailing by 1.2 AICc units. The point estimate is essentially unchanged in this sensitivity, though the data do not distinguish the two specifications and do not rule out mild within-component correlation.

On the full data, 13 of 40 random starts finish within 0.01 log-likelihood units of the best fit, and 27 of 40 report successful convergence; per-replicate original and corrected bootstrap fits are released.

\subsection{Model-comparison results}\label{sec:modelcompresults}

This subsection answers the paper's third question: does anything in the data require a student's midterm score to be related to that same student's final score? The comparison over the candidate set of Section~\ref{sec:modelcomp}, designed to separate dependence from marginal structure, delivers a two-part verdict (Table~\ref{tab:models}): the rigid one-population candidates fail badly, and among the flexible candidates nothing requires dependence. First, the rigid single-population Gaussian candidates receive essentially no support within the candidate set, and there are two distinct reasons why. The two rigid candidates that estimate their between-exam correlation freely settle near zero. The general censored bivariate normal, which contains every jointly normal one-population account, trails the constrained product-of-margins anchor by 19.4 AICc units with a fitted correlation of 0.047; the flexible shared-shift model trails by 36.6 with a fitted correlation of 0.036. Their poor standing therefore cannot be read as a rejection of shared signal alone---it reflects their restrictive single-normal margins as well---though it does mean that the best a jointly normal one-population account can do is to claim that two exams in the same course shared essentially no signal. The calibrated shared-shift model, the one candidate that does impose substantial shared signal, trails by 201.6. All three sit far beyond the point at which the model-selection literature regards a candidate as having essentially no support \citep{burnham2004}. The one-normal-per-margin independence model trails by 17.1 as well, not because independence fails but because a single censored normal cannot describe the ceiling-heavy midterm margin. Flexible margins, not dependence, are what the data demand.

\begin{table}[htbp]

\centering
\small
\begin{tabular*}{\textwidth}{@{\extracolsep{\fill}}llcccc@{}}
\toprule
Model & Dependence & $k$ & $-\log L$ & AICc & $\Delta$AICc \\
\midrule
Product of margins, constrained & none & 7 & 400.1 & 816.3 & 0 \\
Independence, flexible midterm margin & none & 7 & 400.9 & 818.0 & 1.7 \\
Mixture, within-class independence & shared class & 7 & 401.1 & 818.4 & 2.0 \\
Product margins $+$ Gaussian copula & fitted copula correlation & 8 & 399.8 & 818.5 & 2.1 \\
Mixture (main model) & shared class + calibrated & 7 & 401.2 & 818.6 & 2.3 \\
Mixture, free decoupled correlation & shared class + fitted & 8 & 400.4 & 819.8 & 3.4 \\
Independence, one normal per margin & none & 4 & 412.4 & 833.5 & 17.1 \\
General censored bivariate normal & one population, fitted & 5 & 412.3 & 835.7 & 19.4 \\
Shared shift, spread and correlation free & one population, shared & 4 & 422.1 & 852.9 & 36.6 \\
Shared shift, calibrated & one population, shared & 2 & 506.8 & 1017.9 & 201.6 \\
\midrule
Product of margins, unconstrained & none & 7 & 398.9 & 814.1 & excl. \\
\bottomrule
\end{tabular*}
\caption{Does anything in the data require midterm and final scores to be related? No. The top-ranked candidate contains no such link, and a weak link fits nearly as well without being favored. Model comparison over the candidate set of Section~\ref{sec:modelcomp}, which separates dependence from marginal structure, ordered by AICc. All models use the jointly censored likelihood; $k$ counts free parameters, and AICc is the small-sample-corrected Akaike criterion at $n = 59$ (smaller is better; differences are relative support within this candidate set only, not identification of a mechanism or of latent classes). ``Product of margins'' is the product of the main mixture's two marginal distributions; ``shared class'' marks mixture variants whose two scores are dependent through latent class membership even when within-class correlations are zero. The unconstrained product fit (bottom row) chases a censoring artifact---a low-weight final component migrating below the score floor---and is shown as a diagnostic, excluded from the ranking, which the constrained fit anchors. The anchor and the copula fit are boundary-adjacent in the same low-weight final-component mean, and the fine ordering of the leading cluster is sensitive to that regularization (see text and the released bound sensitivity); small $\Delta$AICc values should not be read as a stable ranking.}\label{tab:models}
\end{table}

Second, among the candidates with flexible margins, the data do not require dependence, and weak dependence is not excluded. At the chosen boundary regularization, the top-ranked candidate is the constrained product of the mixture's own margins, which contains no dependence of any kind: its $-\log L$ is 400.1 against the mixture's 401.2 at equal parameter count. The unconstrained product fit reaches 398.9 only by chasing the below-the-floor artifact described in Section~\ref{sec:modelcomp}, and it stays out of the ranking as a diagnostic. The copula candidate addresses weak dependence directly. Allowed to choose any strength of Gaussian-copula dependence on top of the same marginal families, the data choose a dependence parameter of $0.102$, improving the log likelihood by 0.27. The gain is real but small: an added parameter must buy roughly one full log-likelihood unit---slightly more after the small-sample correction---to pay for itself, so the penalized score worsens by 2.1 units. A profile of the dependence parameter, refitting the margins at each fixed value, stays within two log-likelihood units of the best fit between roughly $-0.15$ and $0.35$ (released as a descriptive profile with no formal coverage claim). Read against a substantive threshold, the profile also disfavors dependence of the strength the modeled assistance-free accounts predict: their predicted recorded correlations of 0.4 to 0.7 correspond to latent copula correlations of roughly 0.5 and above, and the profile places even 0.4 some 2.7 log-likelihood units below the best fit, with 0.45 nearly 4 units below---descriptive statements, as above, with no formal coverage claim. The mixture and its variants sit 2.0 to 3.4 units behind the anchor.

Two caveats govern how finely these comparisons can be read, and both trace to the same boundary constraint. Both leading fits---the anchor and the copula---press against the score-scale floor in the same low-weight final-component mean, each at zero to numerical precision. Refitting the anchor under lower bounds of 0.1 to 15 points on that mean raises its AICc by 0.01 to 2.4 units, enough to reorder the very top of the flexible-margin cluster (released sensitivity). The copula was not refit under these bounds, so the sweep speaks to the anchor's own stability, not to the independence-versus-dependence ordering across bounds, about which we make no claim. And AICc is only a rough relative-support score here (Section~\ref{sec:modelcomp}). We therefore draw no conclusion from the fine ordering among the leading flexible-margin candidates and present global independence and weak dependence as a leading cluster rather than declaring a stable winner. What the comparison does establish, within this candidate set and at the chosen regularization, is this: flexible descriptions with no dependence and with weak dependence describe the released pairs about equally well; nothing in the data requires the mixture's latent pair classes; and none of the rigid single-population Gaussian candidates competes with the flexible cluster. The conclusion is not that 96.5 percent of pairs belong to a decoupled class; it is that, at the class level, a model in which knowing a midterm score adds no information about the final describes the data as well as any candidate examined. Dependence is not required and none was detected; the values the data leave open (the copula profile spans roughly $-0.15$ to $0.35$ on its latent scale) include weak-to-moderate association as well as none.

\subsection{Why we do not extrapolate to the full class}\label{sec:classlevel}

The question here is whether these results extend to the full class of 86. They do not. The 27 leavers appear in no released data, so any statement about the full class would require assumptions about them. One might be tempted to go further: multiply the fitted weight by 86 under scenarios for the leavers and compare the result with the professor's public figure of ``at least 50 of 86.'' We do not, because that arithmetic requires reading the mixture's latent classes literally as counts of students, and Sections~\ref{sec:modelcompresults} and \ref{sec:valresults} show that the data neither establish that reading nor calibrate it. What can be said is qualitative. The finishers' pairing shows no detectable dependence. The press-reported facts about the leavers---27 departures after proctoring was announced, roughly 22 of them holding perfect midterms---bear on the selection mechanism, but because the leavers' final scores are unobserved, they say nothing about the leavers' pairing structure in either direction. The released record supports no numerical class-level conclusion, in either direction.

\subsection{Simulation-study results}\label{sec:valresults}

This subsection answers two questions: do simulated assistance-free classes ever look like the released class, and does the fitted index track the true share of decoupled pairs? The answers are essentially never, unless the student-level link between the exams is nearly severed, and no. Table~\ref{tab:valid} reports the simulation study. Recall its design: many simulated classes are built from assistance-free data-generating processes, plus one collaboration-like row-dependence sensitivity, and each class faces two questions. The primary one applies the two direct diagnostics jointly: in the one-sided variant, is the class's correlation no higher than the observed 0.061, with a change spread no smaller than the observed 22.8, and in the symmetric variant, is its \emph{absolute} correlation no larger than 0.061, with the same spread condition? The secondary one asks what index the mixture fits to the class. The joint criteria are the primary endpoint because they involve no optimizer: whatever numerical fitting can or cannot do, these rates cannot be its artifact. Throughout, the reported rates are conditional Monte Carlo results: they describe the examined generators, parameter ranges, attrition rule, and calibration, not the frequency with which such mechanisms would produce the pattern in real courses.

\begin{table}[htbp]

\centering
\footnotesize
\setlength{\tabcolsep}{4pt}
\begin{tabular*}{\textwidth}{@{\extracolsep{\fill}}llccccccc@{}}
\toprule
 & & \multicolumn{4}{c}{Direct diagnostics (no fitting)} & \multicolumn{3}{c}{Mixture index $\hat{q}$} \\
\cmidrule(lr){3-6} \cmidrule(lr){7-9}
Generator & Setting & Mean $r$ & Change SD & Match 1-s. & Match sym. & Mean & Median & $P(\hat{q} {\ge} 0.965)$ \\
\midrule
Shared shift & $\omega = 0$ & 0.60 & 11.3 & 0 & 0 & 0.16 & 0.13 & 0 \\
Heterogeneous shifts & $\omega = 5$ & 0.58 & 12.4 & 0 & 0 & 0.16 & 0.12 & 0 \\
 & $\omega = 10$ & 0.48 & 15.1 & 0 & 0 & 0.36 & 0.35 & 0 \\
 & $\omega = 15$ & 0.42 & 18.4 & 0 & 0 & 0.58 & 0.58 & 0.010 \\
 & $\omega = 20$ & 0.34 & 22.2 & 0.007 & 0.007 & 0.71 & 0.72 & 0.023 \\
Construct overlap & $\lambda = 0.9$ & 0.59 & 18.5 & 0 & 0 & 0.46 & 0.46 & 0 \\
 & $\lambda = 0.75$ & 0.51 & 19.7 & 0 & 0 & 0.56 & 0.57 & 0.003 \\
 & $\lambda = 0.5$ & 0.33 & 20.8 & 0.007 & 0.007 & 0.70 & 0.71 & 0.013 \\
 & $\lambda = 0.25$ & 0.16 & 22.1 & 0.100 & 0.090 & 0.79 & 0.80 & 0.070 \\
 & $\lambda = 0$ & 0.01 & 22.9 & 0.413 & 0.217 & 0.85 & 0.88 & 0.143 \\
Duplicated rows & --- & 0.61 & 11.2 & 0 & 0 & 0.20 & 0.15 & 0 \\
Shared midterm product & --- & 0.43 & 12.6 & 0 & 0 & 0.33 & 0.30 & 0.003 \\
\bottomrule
\end{tabular*}
\caption{Do assistance-free simulations reproduce the released pattern, or calibrate the index? Essentially never; no. Simulated assistance-free classes reproduce the released pattern essentially never, until the student-level link between the exams is nearly severed, and the fitted index does not track the true decoupled share. Simulation study: 300 classes per cell, each analyzed exactly as the released data were. Every design except the last contains no unauthorized assistance (each response is generated from the simulated student's own latent traits); the shared-midterm-product design is a collaboration-like row-dependence sensitivity whose authorization is deliberately unspecified. The primary endpoints involve no model fitting. \emph{Match (1-s.)} is the share of classes at least as extreme as the released data on both diagnostics jointly, in the direction opposite to shared signal ($r \le 0.061$ and change standard deviation $\ge 22.8$; strongly negative correlations count). \emph{Match (sym.)} requires near-zero association ($|r| \le 0.061$, same spread condition). A match rate of 0 means no matches among the 300 simulated classes (resolution 1/300), not a zero probability. The index columns summarize the mixture fit to each class (twenty-four starts; a full thirty-start audit of every replicate applied). All cells use the baseline noise, $\tau = 7$, censoring, observed-midterm attrition, and locations and shifts calibrated to the two observed average scores; the construct-overlap cells additionally calibrate the final construct's spread to the observed final-score spread. $\omega$ is the standard deviation of student-specific genuine shifts; $\lambda$ is the correlation between the constructs the two exams measure, so the $\lambda = 0$ cells are statistically decoupled by construction despite containing no assistance. For comparison, the released data have $r = 0.06$, change standard deviation 22.8, and fitted index 0.965.}\label{tab:valid}
\end{table}

By either criterion, and within the designs examined, classes whose two exams substantially measure the same students essentially never reproduce the released configuration. The joint rate is zero in 300 classes in every heterogeneous-shift cell up to $\omega = 15$---even though at $\omega = 15$ the central 95 percent of students' genuine changes spreads across a band roughly 60 points wide---and a zero here means no matches among the 300 classes simulated, which bounds the cell's rate below roughly 1.3 percent rather than proving it zero. The rate reaches 2 of 300 classes (0.7 percent) at $\omega = 20$, where the simulated change spread matches the observed 22.8, and the one-sided and symmetric counts are identical in all of these cells. The uncertainty of these rates is given here and below as Wilson 95 percent score intervals (0.2 to 2.4 percent for the $\omega = 20$ cell), which quantify only the Monte Carlo uncertainty from simulating 300 classes per cell and are released with the per-cell summaries. The structural reason is visible in the table's correlation column: heterogeneous genuine change weakens rank order without erasing it, so even the most heterogeneous near-parallel classes retain a mean between-exam correlation of 0.34 against the observed 0.06. Duplicated rows and the collaboration-like shared-product design produce no joint matches in 300 classes each.

The construct-overlap cells behave differently, and they are supposed to. These cells weaken the statistical link between the exams by design, and at $\lambda = 0$ the final is statistically independent of the midterm by construction: full statistical decoupling, produced with no assistance of any kind. There the one-sided rate reaches 41 percent (interval 36 to 47 percent) and the symmetric rate 22 percent (interval 17 to 27 percent); the two differ because 59 of the 124 one-sided matches have correlations below $-0.061$, more negative than anything resembling the observed value. At $\lambda = 0.25$ the rates are 10 and 9 percent. High rates in these cells are not false positives: the diagnostics are responding to exactly the near-zero signal those cells were built to contain. What the cells establish is that the statistical pattern does not identify its mechanism. A class whose exams share almost no student-level signal reproduces the released configuration whether the signal was severed by wholesale decoupling of midterms from the students who submitted them, by exams measuring constructs that share essentially no student-level signal, or by some process not simulated here; the released scores cannot say which, and the designs examined are not exhaustive. Considerations outside the scores bear on the construct reading---both exams were written by one instructor for one course, and we are aware of no public account describing the final as covering different material, though the public record is not a complete inventory of the episode---and we present them as context, not as results.

The fitted index tells a second story, about the estimator itself. Across the assistance-free designs, the mean fitted index ranges from 0.16 (the near-parallel designs) to 0.85 (zero construct overlap), and the collaboration-like sensitivity sits at 0.33. An estimator whose average output spans that range when no assistance is present anywhere is not a calibrated estimator of assistance prevalence: its answers do not track a true share. This is the direct reason the paper reports $\hat{q}$ only as a descriptive index of the fitted mixture and attaches no count of students and no conduct reading to it.

The index columns inherit optimizer noise in a way the primary endpoint does not. Although all 3{,}600 fits converge under the twenty-four-start protocol, a full audit refitting every replicate from thirty additional wide starts still improves 887 of the 3{,}600 (25 percent) by more than the 0.01 log-likelihood threshold of Section~\ref{sec:mixfit}, with single-replicate index shifts as large as 0.63; audit corrections are folded into every summary shown, and the index summaries---means and medians as well as threshold rates such as $P(\hat{q} \ge 0.965)$---should be read as approximate. The simulation resolution is 300 classes per cell, so the smallest nonzero rate the table can show is 0.3 percent.

\subsection{Limits of the evidence}\label{sec:cannot}

The final question is what this evidence cannot show. The expanded model comparison shows the data do not establish that the mixture's latent classes exist at all, and the mixture cannot say which pairs, if any, behave as ordinary measurement predicts. These are statements about \emph{patterns}, not people. A student who received no assistance can produce a collapsed pair through a bad day compounding a format change, and a student who received assistance can produce a consistent-looking pair by luck. Along the way to its fit, the algorithm computes a probability for each row (a ``posterior probability''); these have a well-defined meaning under the fitted mixture, but they are not verdicts about students. We report no derived per-student quantity: no posteriors, no classifications, and no row-level influence values, even though anyone with the public data and code can compute them; the anonymized raw score pairs, already public, are the only row-level data anywhere in this work. The reason for declining is not secrecy, since declining hides nothing. The reason is that a ranked list of students would look like evidence about individuals, and it is not.

It is not evidence about individuals for two reasons. A per-pair posterior measures how typical a pair looks under one fitted template rather than the other, under component forms that are themselves assumptions, and the model comparison has already shown that the data do not require those latent classes to exist. The posteriors also carry little discriminating power here: a basic property of the maximum-likelihood fit forces the 59 posteriors to average exactly the fitted weight, 0.965, so most must run high, and whatever variation remains ranks pairs only by their typicality under the two fitted templates, not by anyone's conduct. A per-pair influence value measures how much one pair moves the class-level fit, and the leave-one-out refits bound each pair's effect on the fitted index $\hat{q}$ at 0.018, though other fitted quantities could move differently. No function of two exam scores supplies the causal information that singling out a student would require. Nothing in this paper names any student or establishes misconduct by anyone. The dataset is anonymized, but we cannot stop a reader with outside knowledge of particular scores from trying to link rows to people. That is one more reason to say exactly what these results are: they cannot identify individual misconduct, cannot produce calibrated per-student probabilities, and cannot by themselves justify sanctions against any student.

Individualized adjudication standards, of the kind Brown's academic code committee insisted on \citep{ihe2026, goldstein2026}, are not statistical naivety; they match what distribution-level evidence can and cannot show. The remedies applied in this episode, voiding the midterm for all students and reweighting the final, operated at the level of the assessment rather than of accused individuals. That is the level at which distribution-level evidence speaks. Whether such mid-course changes to an announced grading scheme are sound policy is a separate question that our analysis does not address.

\section{Discussion and conclusions}\label{sec:discussion}

This paper analyzes the publicly released student-level scores from a highly visible academic-integrity episode: a Brown University course whose take-home midterm averaged a reported 96 across the full class and whose proctored final averaged a reported 48.6 among the students who remained (the released pairs for the 59 finishers give 95.7 and 48.8; Table~\ref{tab:verify}), leading the instructor to conclude publicly that most of the class had used generative AI. Public discussion of the episode largely reduced to two positions: the score collapse was self-evident proof of mass cheating, or it was uninterpretable because anxiety, difficulty, selection, and regression to the mean confound any comparison of the two exams. Both positions treat the class averages as the only evidence available, and about averages the second position is correct: each named confound is real, and averages alone cannot distinguish the confounds from unauthorized assistance. The released data, however, contain more than averages. They contain the pairing of each student's two scores, and that pairing supports conclusions sharper than either public position allows. Section~\ref{sec:conclusions} states those conclusions plainly, Section~\ref{sec:limitations} collects the limitations that bound them, and Section~\ref{sec:template} turns the analysis into a protocol for future disputes.

\subsection{Main conclusions}\label{sec:conclusions}

Working only from the released data, and conditional on their accuracy, we answer four questions.

\begin{itemize}
\item \emph{Does the pairing of midterms with finals carry any information?} None that any test detects. The assistance-free explanations, at generous magnitudes, predict a class like Figure~\ref{fig:scatter}(a): scores shifted down, rank order intact, and a correlation of roughly 0.4 to 0.7 between the exams. The released class looks like Figure~\ref{fig:scatter}(b): the correlation is 0.06, individual changes span 104 points, and permutation tests place the observed association squarely inside what random pairing produces. Those tests have more than 90 percent power against association of the strength the explanations predict.
\item \emph{Can the assistance-free explanations reproduce the released pattern?} In the forms modeled, essentially never. Test anxiety, extra difficulty, selection, regression to the mean, and even large heterogeneous ability change all leave rank order visibly intact. Even when we hand the shared-shift account both observed class averages, it produces no joint match in 2{,}000 simulated classes in any configuration that keeps the two exams meaningfully related, and simulated classes with large student-to-student ability changes match at most 0.7 percent of the time.
\item \emph{Does knowing a student's midterm score tell us anything about that student's final score?} Nothing the data require. Among the models compared, and once model complexity is accounted for, one in which a student's midterm carries no information about the same student's final describes the data as well as any candidate examined, a model adding a weak link fits nearly as well without being favored, and the strongly linked accounts fit far worse. The data do not rule out a weak link; they simply do not require any.
\item \emph{What severed the link, and who was involved?} The scores cannot say. In simulation, the released configuration arises essentially only when nearly all student-level signal linking the exams is severed, and more than one mechanism can sever it: wholesale decoupling of midterms from their authors would, but so would a final measuring a nearly unrelated construct, with no assistance involved, and the mechanisms we examined are not exhaustive. The circumstances of the episode bear on the choice among mechanisms; the released scores do not. The fitted mixture index, the weight $\hat{q} = 0.965$ that the model places on its decoupled component, is not a calibrated share of students, so no count of students or papers follows from it, and the evidence cannot establish any individual student's conduct or by itself justify sanctions against anyone.
\end{itemize}

In short, the released scores are strong evidence that students' standings on the midterm did not carry over to the final, and the scores are silent about why.

\subsection{Limitations}\label{sec:limitations}

Several limitations bound our conclusions. In a study of a live public controversy, we must be as transparent about what the analysis cannot show as about what it can.

\begin{itemize}
\item The dataset is as released, not as audited. The score pairs come from a news organization's publication of material the professor submitted to a university committee. They are anonymized and they agree closely with the separately published statistics (Table~\ref{tab:verify}), but the published accounts likely share a common origin and no independent audit of the underlying gradebook exists. All estimates are conditional on the released scores being accurate.
\item The 27 leavers are unobserved. We therefore make no quantitative class-level claims (Section~\ref{sec:classlevel}); every quantitative statement in this paper concerns the 59 finishers. In particular, we do not model selection driven by anticipated final performance or by the revised grade incentives; the two attrition mechanisms we simulate select on shared ability or observed midterms, and richer selection could alter the finishers' pairing pattern in ways the released record cannot check. The simulation rates of Section~\ref{sec:valresults} are computed under those two mechanisms only.
\item The model makes structural assumptions, collected in Section~\ref{sec:assumptions}: bell-curve score components censored at the scale's ends, two exams treated as measures of one construct, independence within decoupled pairs, and a single shared shift for coupled pairs. Boundary scores could instead be treated as structural outcomes in their own right, blank or perfect papers modeled as discrete events rather than censored readings; our blank-final exclusion and boundary-bound sensitivity probe this only partially, and a fully discrete treatment is a natural extension. We probed the assumptions that matter most through the sensitivity sweeps, the model comparison, and the simulation study, and the analysis is fully reproducible, so a critic can replace any assumption and rerun it.
\item The index is imprecise, and it is not a headcount. By the index we mean, here and below, the fitted mixture weight $\hat{q}$, the paper's one-number summary of how far the class departs from ordinary shared-signal measurement. Rebuilding the class repeatedly from its own pairs returns values anywhere from 0.64 to 1.00; assuming the exams are mostly noise pulls the index toward one half (Figure~\ref{fig:calib}); the simulation study shows it does not track the true share of decoupled pairs; and models with no decoupled group at all fit the data just as well. For all of these reasons, we attach no count of students or papers to it. Nor do the data prove that the link between exams is exactly zero: with 59 pairs, the permutation test detects a weak link (a recorded correlation near 0.16) only about a quarter of the time, even though it detects association of the strength the modeled accounts predict more than 90 percent of the time, and a weak link need not be negligible for every purpose. Finally, the fine ranking among the leading models depends on a technical fitting choice (released as a sensitivity analysis), so we draw no conclusion from that ranking.
\item The score data cannot identify the mechanism behind the severed signal (Section~\ref{sec:valresults}). Simulated classes whose two exams share essentially no student-level signal reproduce every summary we examined, whatever real-world process one imagines behind them, including construct divergence, an assistance-free process that produces full statistical decoupling; and the mechanisms we simulated are not an exhaustive list. The considerations that bear on choosing among mechanisms are context from outside the released scores, not statistical results: one instructor wrote both exams for one course's material, and we are aware of no public account describing the final as covering different material.
\item Rows may not be independent: the public record describes near-identical collaborative submissions. Duplicated rows barely move the index, while a shared-midterm-product mechanism inflates it moderately; a full account of within-class dependence is beyond the released data.
\item Decoupling is not a technology finding. In score patterns alone, wholesale use of generative AI and old-fashioned coordinated collusion look the same, so nothing in our results distinguishes them. The stylistic evidence in the public record points to generative AI, but every statistical statement in this paper applies equally to both.
\item Three students scored zero on the final, which may reflect surrender rather than measured proficiency. Excluding them moves the index from 0.965 to 0.963.
\item Individual students are beyond the evidence's reach (Section~\ref{sec:cannot}). The data cannot establish even whether the mixture's coupled class is empty or occupied, let alone which students might belong to it, in either direction.
\item This is one class. The methods transfer to other disputes that pair a contested assessment with a trusted one (Section~\ref{sec:template} gives the protocol), but the numbers are specific to this course, this pair of exams, and this released record.
\end{itemize}

These limitations also define what this paper is and is not. It is a statistical analysis of a released dataset: we test whether the proposed assistance-free explanations can reproduce the data, summarize with the descriptive index how far the released scores depart from the shared-signal structure of ordinary measurement, and mark the point beyond which such evidence cannot go. It establishes no misconduct by any student, and it passes no judgment on any student's character or on the professor's choices of assessment design, mid-course grading changes, or public disclosure. The paper supports the conditional statistical claims just listed, and nothing stronger, in either direction.

\subsection{Using this analysis as a framework}\label{sec:template}

For a future dispute pairing a contested assessment with a trusted one, the protocol is as follows.

\begin{enumerate}
\item \emph{Verify the record.} Check the released student-level data against independently published statistics, state what cannot be verified, and make every conclusion conditional on the record's accuracy (Section~\ref{sec:data}).
\item \emph{Quantify the rival explanations.} Express each proposed assistance-free mechanism as an explicit parameter, anchored in the literature where possible and labeled a generous scenario value where not (Sections~\ref{sec:related} and \ref{sec:assistance-free}). When an explanation cannot be credibly quantified from the available evidence, report that limitation explicitly rather than assigning a convenient value.
\item \emph{Derive the joint signature.} Work out what the rival explanations, at their claimed magnitudes, imply for the joint distribution of paired scores---here, rank order preserved and a roughly common shift---so that the dispute turns on a testable pattern rather than on averages (Sections~\ref{sec:problem} and \ref{sec:assistance-free}).
\item \emph{Test model-free first.} Permutation tests of the pairing and summary diagnostics compared against simulated benchmark grids use few assumptions (Section~\ref{sec:modelfree}). If the pairing shows strong association and the diagnostics sit inside the benchmark range, the rival accounts survive, and the analysis can stop here.
\item \emph{Give the rival account its best case.} Concede the rival account whatever cannot be denied. Here that means both observed averages and the full examined range of noise levels, with a companion fit that lets the noise float freely; then test only the structure that remains (the mean-matched design and free-noise fit of Section~\ref{sec:modelfree}). This step is the general answer to disputes over assumed effect sizes.
\item \emph{Where feasible, fit a model that nests the principal quantified rival account}, so that data needing no decoupled component can say so by returning an index near zero. Compare it against alternatives built to separate dependence from marginal flexibility (Sections~\ref{sec:mixture} and \ref{sec:modelcomp}).
\item \emph{Stress-test the analysis on simulated classes where the truth is known}, including generators outside the fitted family and assistance-free generators that reproduce the suspicious pattern; check whether the fitted index is calibrated as a share of students, and do not assume that it is (Sections~\ref{sec:validation} and \ref{sec:valresults}).
\item \emph{State the limits as findings.} What the evidence cannot show---individual conduct, a prevalence, the mechanism behind a severed signal---is established by steps 6 and 7, not merely conceded (Sections~\ref{sec:classlevel}, \ref{sec:valresults}, and \ref{sec:cannot}).
\end{enumerate}

The protocol requires three ingredients: per-student paired scores on one contested and one trusted assessment, both intended to measure the same construct; an independently published record to verify against; and enough students for permutation tests to resolve. A bounded score scale is not itself a requirement: where scores are bounded, as here, the censoring treatment handles the ceilings and floors, and where they are not, that machinery is simply unnecessary. When a condition fails, part of the protocol survives. Without a trusted comparison assessment, only each exam's own distribution can be examined and the pairing analyses are unavailable; with very few students, the permutation tests lose resolution first. Some elements transfer directly: the ordering from model-free to model-based, the nesting of the rival account, the mean-matched concession design, the separation of dependence from marginal flexibility, the simulation study, and the guardrails of step 8. Others must be adapted to the data at hand: the calibration constants, the candidate set, and the treatment of boundary scores. Still others are specific to this episode: the scenario magnitudes, the attrition mechanisms, and the boundary regularization that this dataset's ceiling-heavy margins forced. Steps 1 through 4 require only a scatterplot, correlations, permutation tests, and simulation from explicit generators, all within reach of a quantitatively trained education researcher; steps 5 through 7 involve censored likelihoods and model comparison near parameter-space boundaries, where statistical collaboration is warranted. When the conditions fail badly---provenance that cannot be verified, assessments with no case for comparability, or fits pinned to parameter-space boundaries with unstable rankings---the defensible stopping point is the model-free diagnostics, reported without scalar mixture summaries and, where comparability itself is in doubt, read only as descriptions of the observed pairing rather than as evidence that either assessment failed to preserve an expected ordering.

Score-level evidence can be far stronger than the current public debate assumes. Here, a permutation test, a correlation, and a mixture model disciplined by a simulation study were enough to show that the pairing of two exams in one course carried no detectable information. The assistance-free accounts we examined almost never reproduce that pattern while substantial cross-exam signal remains. Evidence of this kind speaks to the assessment: under the stated assumptions, it can show that an assessment did not preserve any detectable ordering of the students it was meant to measure. Whether and how an instructor should act on that conclusion during a semester is a separate question of academic policy that our analysis does not address, and nothing in it endorses or criticizes the particular remedies used in this episode. At the same time, what the evidence cannot do, even at its strongest, is name individuals. If some students' midterms were genuinely their own work, those students are invisibly mixed among the collapsed scores, and a policy that punished by score pattern would punish some of them. As generative AI makes disputes of this kind routine, institutions would do well to make the assistance-free explanations quantitative enough to be tested, to take the aggregate evidence seriously when those explanations fail, and to respect the boundary between characterizing a distribution and accusing a person.

\backmatter

\section*{Declarations}

\bmhead{Ethics approval}

\noindent
This study analyzed publicly released, de-identified secondary data. No new
data were collected from human participants, no interaction with human
subjects occurred, and no individually identifying information was accessible
to the authors at any point. Under United States federal regulations at 45 CFR
46.102(e), research using publicly available information that is not
individually identifiable does not meet the definition of human subject
research.

\bmhead{Availability of data and materials}

\noindent
A replication package is available at \url{ https://doi.org/10.5281/zenodo.21693700 }
\citep{topaz2026archive}. It includes all code used to generate the results
reported here, together with a copy of the raw data, which we extracted on 16
July 2026 from the publicly released interactive chart accompanying Inside
Higher Ed's 8 July 2026 report of this episode \citep{ihe2026,
serranodata2026}.

\bmhead{Competing interests}

\noindent
The authors declare that they have no competing interests.

\bmhead{Funding}

\noindent
This research received no funding.

\bmhead{Authors' contributions}

\noindent
CMT: conceptualization, methodology, software, formal analysis, validation, visualization, writing --- original draft, writing --- review and editing. UB: methodology, writing --- review and editing. Both authors read and approved the final manuscript.

\bmhead{Use of generative AI}

\noindent
We disclose our own use of generative AI because a paper about unauthorized AI
assistance in assessment owes its readers an account of its own practice. We
used Claude (Anthropic), primarily the Fable 5 model, as an assistant in four
ways: assisting in code development; searching for and assembling candidate
literature; editing prose; and checking the manuscript for internal
consistency and stylistic uniformity. The research questions, the statistical
design, the choice of models and diagnostics, the interpretation of every
result, and the argument the paper makes are the authors' own. No text entered
the manuscript without being read and revised by an author, no numerical
result was accepted without being reproduced by the released pipeline, and no
reference was accepted without being checked against the original source. No
large language model is an author of this work; the authors take full
responsibility for every claim, number, citation, and line of code in the
article and in the released repository.

\bibliography{refs}

\end{document}